\documentclass[twocolumn]{aastex7}

\usepackage{graphicx}   
\usepackage{multirow}
\usepackage{amsmath}
\usepackage{lineno}
\usepackage{subcaption}
\usepackage{float}
\usepackage{afterpage}
\usepackage{booktabs}
\usepackage{siunitx}
\usepackage{multirow} 
\usepackage{arydshln}

\graphicspath{{./}{figures/}}

\shorttitle{width and profiles of cosmic filaments}
\shortauthors{Yang, Zhu, Yu, Mo, Zheng \& Feng.}

\begin{document}

\title{On the width and profiles of cosmic filaments}

\author{Qi-Rui Yang}
\email{yangqr8@mail2.sysu.edu.cn}
\affil{School of Physics and Astronomy, Sun Yat-Sen University, Zhuhai campus, No. 2, Daxue Road \\
Zhuhai, Guangdong, 519082, China}
\affil{CSST Science Center for the Guangdong-Hong Kong-Macau Greater Bay Area, Daxue Road 2, 519082, Zhuhai, China}

\author[0000-0002-1189-2855]{Weishan Zhu}
\email{zhuwshan5@mail.sysu.edu.cn}
\affil{School of Physics and Astronomy, Sun Yat-Sen University, Zhuhai campus, No. 2, Daxue Road \\
Zhuhai, Guangdong, 519082, China}
\affil{CSST Science Center for the Guangdong-Hong Kong-Macau Greater Bay Area, Daxue Road 2, 519082, Zhuhai, China}

\author[0009-0001-6368-2833]{Guan-Yao Yu}
\email{yugy9@mail.sysu.edu.cn}
\affil{School of Physics and Astronomy, Sun Yat-Sen University, Zhuhai campus, No. 2, Daxue Road \\
Zhuhai, Guangdong, 519082, China}
\affil{CSST Science Center for the Guangdong-Hong Kong-Macau Greater Bay Area, Daxue Road 2, 519082, Zhuhai, China}

\author{Jian-Feng Mo}
\email{mojf3@mail2.sysu.edu.cn}
\affil{School of Physics and Astronomy, Sun Yat-Sen University, Zhuhai campus, No. 2, Daxue Road \\
Zhuhai, Guangdong, 519082, China}
\affil{CSST Science Center for the Guangdong-Hong Kong-Macau Greater Bay Area, Daxue Road 2, 519082, Zhuhai, China}

\author{Yi Zheng}
\email{zhengyi27@mail.sysu.edu.cn}
\affil{School of Physics and Astronomy, Sun Yat-Sen University, Zhuhai campus, No. 2, Daxue Road \\
Zhuhai, Guangdong, 519082, China}
\affil{CSST Science Center for the Guangdong-Hong Kong-Macau Greater Bay Area, Daxue Road 2, 519082, Zhuhai, China}

\author{Long-Long Feng}
\email{flonglong@mail.sysu.edu.cn}
\affil{School of Physics and Astronomy, Sun Yat-Sen University, Zhuhai campus, No. 2, Daxue Road \\
Zhuhai, Guangdong, 519082, China}
\affil{CSST Science Center for the Guangdong-Hong Kong-Macau Greater Bay Area, Daxue Road 2, 519082, Zhuhai, China}

\correspondingauthor{Weishan Zhu}
\email{zhuwshan5@mail.sysu.edu.cn}


\begin{abstract}
We investigated the widths and profiles of cosmic filaments using the IllustrisTNG simulations. Filaments were identified with DisPerSE, using galaxy samples in simulations as input. Since the width of an individual filament can vary significantly along its spine, we divided each filament into segments with lengths between $1.5\,h^{-1}\,\mathrm{Mpc}$ and $2.5\,h^{-1}\,\mathrm{Mpc}$ and measure their properties. The typical width of these filament segments increases gradually from approximately $0.3\,\mathrm{Mpc}$ at redshift $z = 2.0$ to about $1.0-1.5\,\mathrm{Mpc}$ at $z = 0.0$. We find that the segment width correlates nearly linearly with the linear halo mass density, consistent with previous studies. A similar linear relation is observed between the segment width and the linear stellar mass density, providing a potential estimator for filament width. Furthermore, the density profiles of filaments with different widths exhibit self-similarity and can be described by a unified formula akin to the isothermal $\beta$-model. For segments with a given width, the rescaled density profiles show only mild evolution from $z = 2.0$ to $z = 0.0$. Within the filament width, the gas temperature decreases slowly from the center to the boundary, with thicker filaments generally containing hotter gas than thinner ones. These trends in filament width, density, and thermal profiles are consistently observed across the TNG50, TNG100, and TNG300 simulations, and align well with results from earlier studies. We briefly discuss the potential implications and applications of our findings.

\end{abstract}



\section{Introduction} \label{sec:intro}

The large-scale distribution of galaxies forms a complex, web-like structure known as the cosmic web, comprising clusters (or nodes), filaments, walls (or sheets), and voids \citep[e.g.]{1986ApJ...302L...1D,2003astro.ph..6581C,2014MNRAS.438..177A,2014MNRAS.438.3465T,laigle2025euclid}. This intricate pattern was first predicted by theoretical studies, which attributed its emergence to the anisotropic gravitational collapse of matter in the expanding universe (e.g. \citealt{1970A&A.....5...84Z,1996Natur.380..603B,2008LNP...740..335V}). Over the past two decades, the structure and properties of the cosmic web have been extensively investigated through both observations and numerical simulations (e.g. \citealt{2005MNRAS.359..272C,2007A&A...474..315A,2009MNRAS.396.1815F,2010ApJ...723..364A,2010MNRAS.408.2163A,2012MNRAS.425.2049H,2014MNRAS.441.2923C,2017ApJ...837...16D,2017ApJ...838...21Z,2018MNRAS.473.1195L,2019MNRAS.485.4743S}).

Cosmic filaments contain the largest fraction of matter among all cosmic web structures at redshifts $z \lesssim 2$ (e.g. \citealt{2010MNRAS.408.2163A,2014MNRAS.441.2923C, 2017ApJ...838...21Z}) and are believed to host the majority of the missing baryons in the low-redshift universe (e.g. \citealt{1998ApJ...503..518F,1999ApJ...514....1C,2012ApJ...759...23S,2019MNRAS.486.3766M}). Observational evidence for ionized baryons in filaments has been reported through both X-ray detections and measurements of the thermal Sunyaev-Zel'dovich (SZ) effect (e.g. \citealt{2002ApJ...572L.127F,2009ApJ...699.1765B,2019A&A...621A..88N,2019A&A...624A..48D,2019MNRAS.483..223T,2020A&A...643L...2T,2022A&A...667A.161T}).

In addition, numerous studies have indicated that cosmic filaments significantly influence the properties of nearby halos and galaxies (e.g. \citealt{2007MNRAS.375..489H,2017MNRAS.466.1880C,2017A&A...600L...6K,2018MNRAS.474..547K,2018MNRAS.476.4877M,winkel2021imprint,castignani2022virgo,song2021beyond,2022A&A...658A.113M,bulichi2024galaxy,laigle2025euclid}). However, other studies suggest that the impact of filaments on galaxy properties may be minor or moderate when controlling for multiple factors such as, local overdensity, and halo mass (e.g. \citealt{kuutma2017voids,hoosain2024effect,o2024effect,hasan2024filaments}). Recent work by \citet{2025arXiv250401245Y} further demonstrates that, once local overdensity is taken into account, the differences among central galaxies across various cosmic web environments largely disappear. In contrast, residual differences remain for satellite galaxies, particularly between those residing in the field and those in filament-dominated regions, even after controlling for stellar mass, halo mass, and local overdensity.

A comprehensive understanding of the properties of cosmic filaments is essential for addressing relevant key questions in astrophysics and cosmology, such as detecting the missing baryons, clarifying the influence of filaments on galaxy evolution, and using filaments as probes of cosmological models and the nature of gravity (e.g. \citealt{2018MNRAS.479..973C,2018A&A...619A.122H}). However, significant discrepancies remain in the literature regarding filament properties, including their lengths, widths, matter density profiles, and gas temperature distributions (e.g. \citealt{2005MNRAS.359..272C,2010MNRAS.408.2163A,2014MNRAS.441.2923C,2019MNRAS.486..981G,2020A&A...641A.173G,2021ApJ...920....2Z,2024MNRAS.52711256L,2024A&A...684A..63G}). These inconsistencies largely stem from variations in filament samples, as well as differences in the definitions, classification criteria, and measurement methodologies employed across studies. Without robust and consistent estimates of filament properties, it remains challenging to resolve critical issues such as the role of the cosmic web in shaping galaxy properties and the localization of the universe’s missing baryons.

The width of cosmic filaments is a key property, yet its accurate measurement remains a significant challenge. Several studies have estimated filament radii based on density profiles or the radial distribution of halos and galaxies, using both simulations and observational data (e.g. \citealt{2005MNRAS.359..272C,2010MNRAS.408.2163A,2010MNRAS.409..156B,2010MNRAS.407.1449G,2014MNRAS.441.2923C,2014MNRAS.438.3465T,2020A&A...638A..75B,2021ApJ...920....2Z,2023MNRAS.525.4079Z,2024MNRAS.532.4604W}). However, reported values for filament widths vary significantly—ranging from 0.5 to 8.4 Mpc—primarily due to differences in filament selection, classification methods, and boundary definitions. Observational estimates are affected by biases due to incomplete galaxy samples, especially at intermediate and high redshifts. In contrast, simulations—particularly those utilizing the underlying density field—can provide more complete and less biased measurements. However, caution is necessary when comparing results across different simulations or between simulations and observations, as variations in filament identification methods and boundary definitions can introduce significant inconsistencies.

Moreover, the width of a single cosmic filament can vary significantly along its spine (e.g. \citealt{2014MNRAS.441.2923C,2015MNRAS.453.1164G,2017ApJ...838...21Z,2019MNRAS.486..981G}). As a result, the local width, defined as the width of a filament segment over a length of approximately 2–3 Mpc (e.g. \citealt{2014MNRAS.441.2923C,2021ApJ...920....2Z}), could offer a more precise characterization. Using a high-resolution cosmological hydrodynamic simulation, \cite{2024ApJ...967..141Z} found a correlation between the local filament width and the enclosed dark matter halo mass per unit length. While this relationship presents a promising approach to estimating local filament widths, it has yet to be tested in other simulations. Furthermore, its application to observational galaxy samples remains challenging, as accurately measuring the halo masses of individual galaxies is non-trivial. Therefore, it is important to further validate this correlation using a variety of simulations and to explore whether it can be extended to observable galaxy properties that are more readily accessible in surveys.

The matter density and temperature profiles of cosmic filaments are also fundamental properties, yet they display significant variation across different studies ( \citealt{2010MNRAS.408.2163A,2014MNRAS.441.2923C,2019MNRAS.486..981G,2021A&A...649A.117G,2021A&A...646A.156T}).This discrepancy may partly arise from differences in simulation methodologies—such as the gravity and hydrodynamical solvers—and variations in subgrid physics, including the treatment of stellar and AGN feedback. Additionally, the filament samples used across different studies often vary in width, which can further contribute to discrepancies in the resulting density and temperature profiles. \cite{2021ApJ...920....2Z} found that density profiles across filaments of varying widths exhibit self-similarity when normalized by their respective widths and can be described by an isothermal single-beta model. Additionally, the study showed that thicker filaments tend to contain hotter gas, a trend consistent with the findings of \cite{2021A&A...649A.117G} and \cite{2021A&A...646A.156T}. More recently, while this work was in preparation, \citet{2025arXiv250206484B} also reported the gas in thicker filaments are hotter.

In this study, we investigate the widths and internal profiles of cosmic filaments using the IllustrisTNG simulations. We explore potential estimators for the local width of filaments, including the linear halo mass density and linear stellar mass density. We then probe the density and temperature profiles of filament segments with different widths. This paper is organized as follows. Section 2 provides a brief overview of the cosmological simulation and the numerical methods used to identify and segment filaments, as well as to measure their  widths and profiles. In Section 3, we present the distribution of segment width, the correlations between filament segment width and the linear halo/stellar mass densities at $z=0$. Section 4 details the density and temperature profiles of filament segments at $z=0$. Results at higher redshifts are reported in Section 5 Finally, we summarize our findings in Section 6.  

\begin{figure}[htb]
\begin{centering}
\hspace{-0.5cm}
\includegraphics[width=0.33\textwidth, clip]{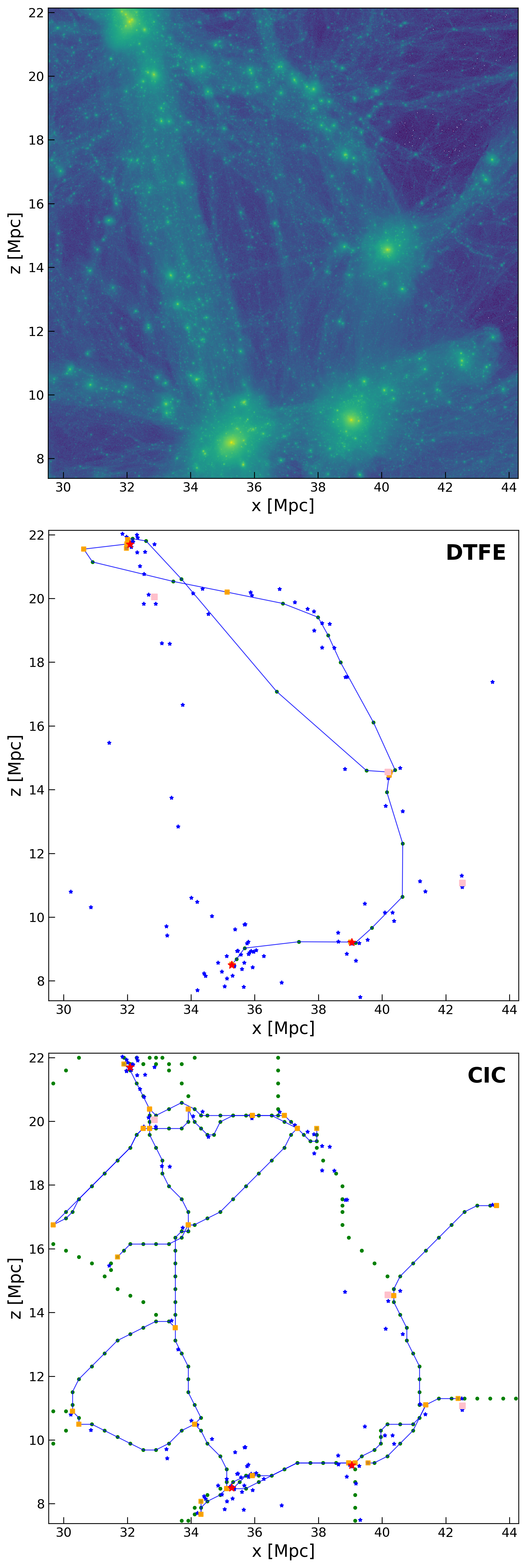}
\caption{The dark matter distribution (top) and filaments identified by DisPerSE using galaxy density fields derived from DTFE (middle) and CIC (bottom) in a subbox of thickness 5 Mpc from TNG50. Blue lines represent filaments, green dots indicate segment (sampling) points, orange squares indicate filament nodes(endpoints).red stars mark halos with masses greater than $10^{13}\,\text{M}_{\odot}$, pink squares denote halos with mass exceeding $10^{12}\,\text{M}_{\odot}$., blue stars are input tracers (galaxies with $\text{M}_{\star} > 10^9$).}
\label{fig:dtfe_cic}
\end{centering}
\end{figure}

\section{Methodology} \label{sec:method}

\subsection{IllustrisTNG Simulations } 
In this study, we make use of the TNG50, TNG100, and TNG300 simulations from the IllustrisTNG project (\citealt{2018MNRAS.475..624N,2018MNRAS.475..648P,2018MNRAS.475..676S,2018MNRAS.477.1206N,2018MNRAS.480.5113M}). These large-scale cosmological simulations follow the co-evolution of dark matter, baryonic gas, stars, and supermassive black holes, and are performed using the moving-mesh code AREPO (\citealt{springel2010pur}). They incorporate key baryonic physics processes, including radiative cooling, star formation, chemical enrichment, stellar feedback, and active galactic nucleus (AGN) feedback. The simulations trace the formation and evolution of cosmic structures from redshift $z = 127$ to $z = 0$. A standard $\Lambda$CDM cosmology is assumed, based on parameters from the Planck 2016 results (\citealt{ade2016planck}): $h = 0.6774$, $\Omega_{\mathrm{m}} = 0.3089$, $\Omega_{\Lambda} = 0.6911$, $\Omega_{\mathrm{b}} = 0.0486$, $\sigma_8 = 0.8159$, and $n_s = 0.9667$.

The primary differences among TNG50, TNG100, and TNG300 lie in their simulation volumes and resolutions. Specifically, TNG50, TNG100, and TNG300 cover comoving volumes of $(51.7\,\text{Mpc})^3$, $(110.7\,\text{Mpc})^3$, and $(302.6\,\text{Mpc})^3$, respectively. The dark matter particle masses are $4.5 \times 10^5\,\text{M}_{\odot}$ in TNG50, $7.5 \times 10^6\,\text{M}_{\odot}$ in TNG100, and $5.9 \times 10^7\,\text{M}_{\odot}$ in TNG300. The corresponding baryonic mass resolutions are $8.5 \times 10^4\,\text{M}_{\odot}$, $1.4 \times 10^6\,\text{M}_{\odot}$, and $1.1 \times 10^7\,\text{M}_{\odot}$, respectively. TNG50 achieves the highest spatial resolution (softening length), reaching $\sim$100–140 pc for gas in the interstellar medium and 20–80 pc in star-forming regions at $z=0$. In comparison, TNG100 reaches resolutions of approximately 190 pc and 140 pc in these respective environments, while TNG300 achieves around 1 kpc and 500–800 pc. The spatial resolution for dark matter is roughly four times coarser than that for gas. Notably, the resolution for dark matter particles remains constant between redshift $0$ and $1$, and then scales as $1/(1+z)$ at $z>1$. In contrast, the resolution for gas is adaptive and set to 2.5 times the comoving cell radius. By analyzing and comparing results across these three simulations in snapshots 99 (z=0), 67 (z=0.5), 50(z=1.0), and 33 (z=2.0), we can assess how simulation volume and resolution influence the measured properties of cosmic filaments.

\begin{figure*}[htbp]
\begin{centering}
\hspace{-0.0cm}
\includegraphics[width=0.75\textwidth, clip]{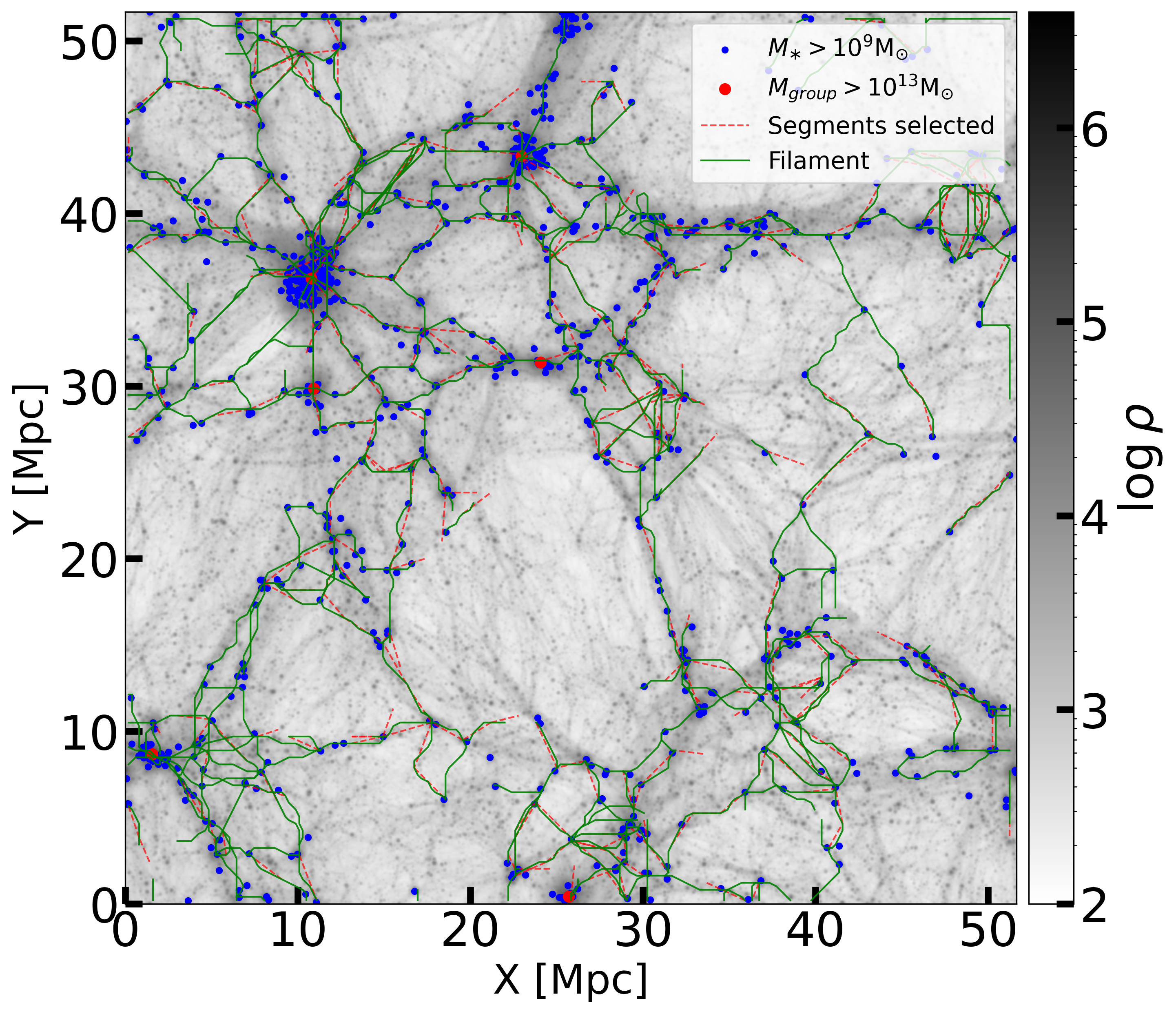}
\caption{Filaments (green lines) within a slice in TNG50 found by DisPerSE using cic mass weighted grid density field. Blue filled circles are galaxies with stellar mass greater than $10^9\,M_{\odot}$. Red filled circles are groups with total mass greater than $10^{13}\,M_{\odot}$. The background gray plot shows the dark matter density distribution. Red dashed lines indicate relatively straight segments after segmentation.  }
\label{fig:filament_in_slice}
\end{centering}
\end{figure*}

\subsection{Filaments identification and segmentation}

We employ the widely used tool Discrete Persistent Structures Extractor (\texttt{DisPerSE}; \citealt{sousbie2011persistent, sousbieet2011persistent}) to identify the cosmic filamentary structures in the IllustrisTNG simulations. The procedure is as follows: first, we compute the density field using galaxies (subhalos) with stellar masses $M_{\ast} > 10^9\,\text{M}_{\odot}$, consistent with the threshold adopted in many observational studies that use galaxies as tracers to identify filaments (\citealt{2014MNRAS.438.3465T, donnan2022role, o2024effect}). 

It is worth noting that, instead of using the delaunay$\_$nD function, based on the Delaunay Tessellation Field Estimator (DTFE, \citealt{2000A&A...363L..29S}), to estimate the density field, we adopt a mass-weighted Cloud-in-Cell (CIC) algorithm. This choice is motivated by the relatively limited number of galaxies (subhalos) in the simulations compared to the number of dark matter particles. The DTFE method estimates the density field by computing the volume of each simplex (tetrahedron) in the tessellation and distributing the density equally among its four vertices. As a result, it requires a sufficiently large number of input points to produce an accurate and stable estimate. To evaluate the performance of both methods, we applied the CIC and DTFE algorithms to the TNG50 simulation, using galaxies with stellar masses above $10^9 \text{M}_{\odot}$ as input. We then followed the same filament extraction procedures for both cases. As shown by Figure \ref{fig:dtfe_cic}, we found that the filaments identified using the mass-weighted CIC density field aligned more closely with the actual matter distribution, supporting the use of the CIC method in our analysis.

The density field obtained via the CIC algorithm is smoothed before being used as input for DisPerSE. The grid resolution for the CIC computation and the Gaussian smoothing scale are set to 0.404$\,\mathrm{cMpc}$ for TNG50, 0.433$\,\mathrm{cMpc}$ for TNG100, and 0.590$\,\mathrm{cMpc}$ for TNG300. Note that these smoothed density fields are used solely to identify the filament skeleton. The original, unsmoothed distributions of dark matter and gas particles will be used for computing density and temperature profiles and measuring widths in the analyses that follow. Filaments are identified using the mse function in DisPerSE, which relies on two fundamental concepts: critical points, where the gradient of the density field vanishes, and integral lines, which are curves tangent to the gradient vector at each point in the field. In our three-dimensional samples, four types of critical points are identified—minima, saddle-1, saddle-2, and maxima—indexed from zero to three. We treat bifurcation points as nodes, which can be interpreted as cluster centers. Filaments are defined as lines that originate from a maximum, pass through a saddle-2 point, and terminate at another maximum.

The influence of various parameter choices in DisPerSE has been investigated in previous studies (e.g., \citealt{Duckworth_2019,Duckworth_2020}). In this work, we adopt a persistence threshold of -Nsig 4. According to the official DisPerSE manual, the -Nsig parameter is primarily intended for use with density fields estimated via the Delaunay Tessellation Field Estimator (DTFE), while the -Cut parameter is recommended for other types of density fields. Nevertheless, we find that it works well with -Nsig 4 for our density fields calculated by CIC. In contrast, determining an optimal value with the -Cut parameter is significantly more time-consuming. Therefore, we use -Nsig 4 in this work. 

As an illustration, Figure \ref{fig:filament_in_slice} displays the filaments identified within a 10 Mpc-thick slice of the TNG50 simulation at redshift $z=0$. Blue dots represent galaxies with stellar mass $M_{\ast}>10^9\, \text{M}_{\odot}$, while red dots mark galaxy groups with total mass exceeding $10^{13}\, \text{M}_{\odot}$. The solid green lines correspond to the filamentary structures extracted using \texttt{DisPerSE}. The background grayscale map shows the dark matter density field. As shown, the identified filaments effectively trace both the large-scale cosmic web outlined by the galaxy distribution and the underlying matter density field trace the distribution of matter. 

The filament samples identified by \texttt{DisPerSe} span a wide range of lengths, from approximately $1.5\,\mathrm{Mpc}$ to over $15\,\mathrm{Mpc}$ across the three simulations. The width (or thickness) of individual filaments can vary significantly along their spines, and filaments themselves are often curved. To study the local width and associated profiles more accurately, we divide each filament into relatively straight segments with lengths between $1.5\text{Mpc/h}$ ($2.2\text{Mpc}$) and $2.5\text{Mpc/h}$ ($3.7\text{Mpc}$). The segmentation procedure is detailed in the Appendix. The red dashed lines in Figure \ref{fig:filament_in_slice} indicate the relatively straight segments after segmentation, which will be further probed in the following Sections. As a result, we obtain 1,009, 8,482, and 134,361 filament segments at $z=0$ in TNG50-1, TNG100-1, and TNG300-1, respectively.  

\subsection{width/thickness and profiles}

\begin{figure*}[htb]
\begin{centering}
\hspace{-0.0cm}
\includegraphics[width=1.0\textwidth, clip]{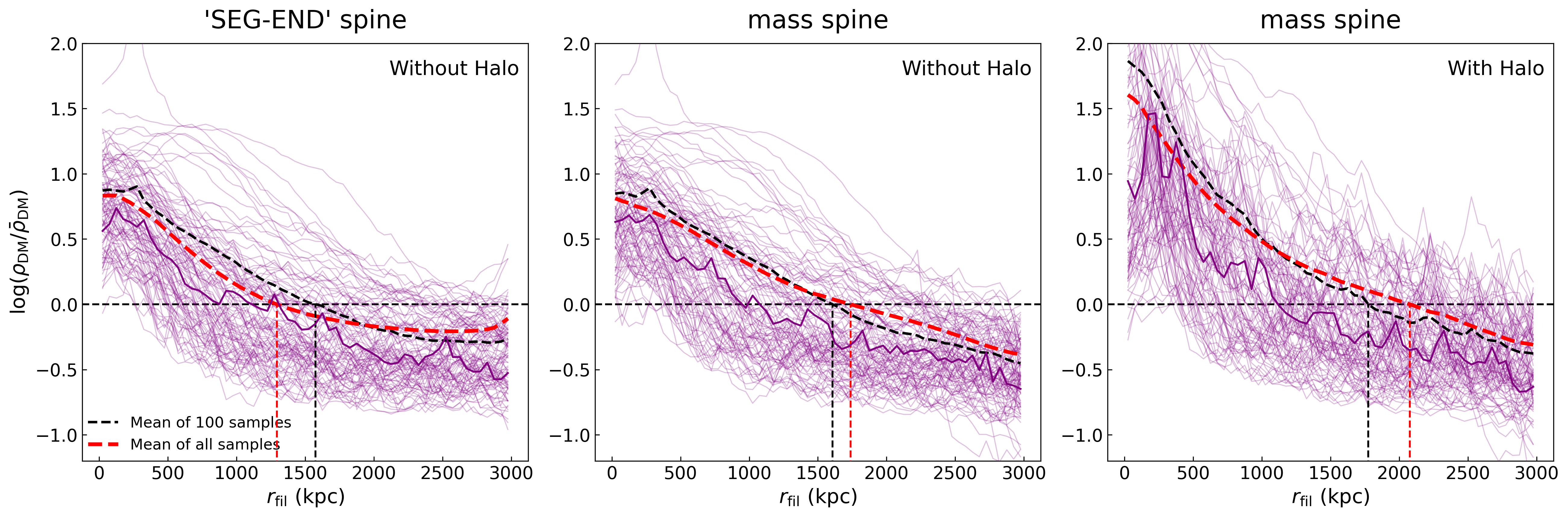}
\caption{Dark matter density profiles of randomly selected 100 filament segments in TNG100 at redshift $z=0$. The left panel shows results using the SEG-END spine as the axis, excluding matter within halos more massive than $10^{10},\mathrm{M_{\odot}}$ (see main text for details). The middle panel presents results using the mass spine as the axis, also excluding halo matter. The right panel displays results based on the mass spine, but including contributions from matter within halos. The thick black and red dashed lines show the mean profiles of the 100 selected segments and all TNG100 segments, respectively. Thin vertical dashed lines indicate where the mean density profiles drop to the cosmic mean.}
\label{fig:sample_profiles}
\end{centering}
\end{figure*}

We measure the width, as well as the density and temperature profiles of filament segments in our samples using the original positions of gas and dark matter particles from the TNG simulations around each segment, by the following procedure. Note that no smoothing is applied when computing the profiles for the segments. The first step is to define the segment spine or axis, for which we explore two methods. In the first approach, referred to as the `SEG-END' spine, the spine is simply defined as the straight line connecting the two endpoints of a filament segment. In the second approach, we compute the center of mass within a cylindrical region of radius $2\,\rm{Mpc}/h$ (2.95 Mpc), where the cylinder axis is aligned along the `SEG-END' spine, similar to some previous studies(\citealt{2021MNRAS502351R,2024MNRAS.52711256L}). The upper and lower surfaces of the cylinder are defined as planes perpendicular to the `SEG-END' spine that pass through the two endpoints. We then define the mass spine as the line that runs parallel to the `SEG-END' spine and passes through this center of mass. As illustrated in Figure \ref{fig:two_spines}, these two spine definitions can diverge for several reasons: galaxy distribution do not perfectly trace the underlying matter distribution, filament segments may be curved, or the matter distribution may lack axial symmetry. Note that, we adopt a threshold of $2\,\rm{Mpc}/h$ primarily here to avoid contamination from nearby filaments. However, a small fraction of segments can in fact be thicker than this threshold. 

We approximate each filament segment as a cylindrical structure and define its width (or thickness) as the cylinder’s radius. To determine this, we compute the radial density profiles of gas and dark matter. Using either the stellar or mass spine as the central axis, we calculate the mass of gas cells or dark matter particles within concentric cylindrical shells and normalize it by the volume of each shell to obtain the density profile.
The filament width is then defined as the radius at which the gas or dark matter density drops to the cosmic mean density of the respective component. We denote the radii determined from the dark matter and gas density profiles as $R_{fil,dm}$ and $R_{fil,gas}$, respectively.

It is important to note that the radius can be defined in two ways: one includes all matter within the segment, including that bound within halos, while the other excludes halo-bound matter when computing the density profile. Halos with masses above $10^{10}\, M_{\odot}$ account for approximately $90\%$ of the total halo mass. To improve computational efficiency, we will explore the profile both before and after removing the matter within halos that are more massive than $10^{10}\, M_{\odot}$ and use these to determine the width/radius. We find that the width given by the former method usually is about $1.1-1.3$ times of the latter. We refer to the former as the extended width (radius), whereas the latter focuses on the diffuse matter and serves as our default definition of filament width. Moreover, we focus on filament segments with a width between $10^{2.5}\, \rm{kpc}$ and $10^{3.5}\, \rm{kpc}$ in this work because following reasons. First, it is relatively difficult to accurately measure the properties of filaments thinner than $\sim 0.3\, \rm{Mpc}$. Second, while a small fraction of segments have widths exceeding $\sim 3\, \rm{Mpc}$, their density profiles at large radii are often contaminated by nearby filaments, making them less suitable for our analysis.

As an illustration, Figure~\ref{fig:sample_profiles} shows the dark matter density profiles of 100 filament segments from the TNG100 simulation. The left and middle panels present profiles computed using the `SEG-END' spine and the mass spine, respectively, excluding matter bound to halos more massive than $10^{10}\,\text{M}_{\odot}$. The right panel displays profiles based on the mass spine, but including contributions from halo matter. The density profiles of individual segments exhibit significant fluctuations, particularly when halo matter is included. The black dashed lines represent the mean profiles of the 100 sampled segments, while the red dashed lines indicate the mean profiles of all segments in TNG100. Profiles computed using the mass spine are generally more extended than those based on the SEG-END spine, as shown by the red dashed lines. Moreover, the stellar component and galaxy positions, on which the SEG-END spine is based, are biased tracers of the total matter distribution. Consequently, profiles computed using the SEG-END spine are unlikely to accurately capture the underlying mass structure of filaments. For this reason, we adopt profiles centered on the mass spine for our analysis. The corresponding widths derived from these profiles are used as the default in the subsequent sections.

\begin{figure*}[htb]
\begin{centering}
\hspace{-0.0cm}
\includegraphics[width=1.0\textwidth, clip]{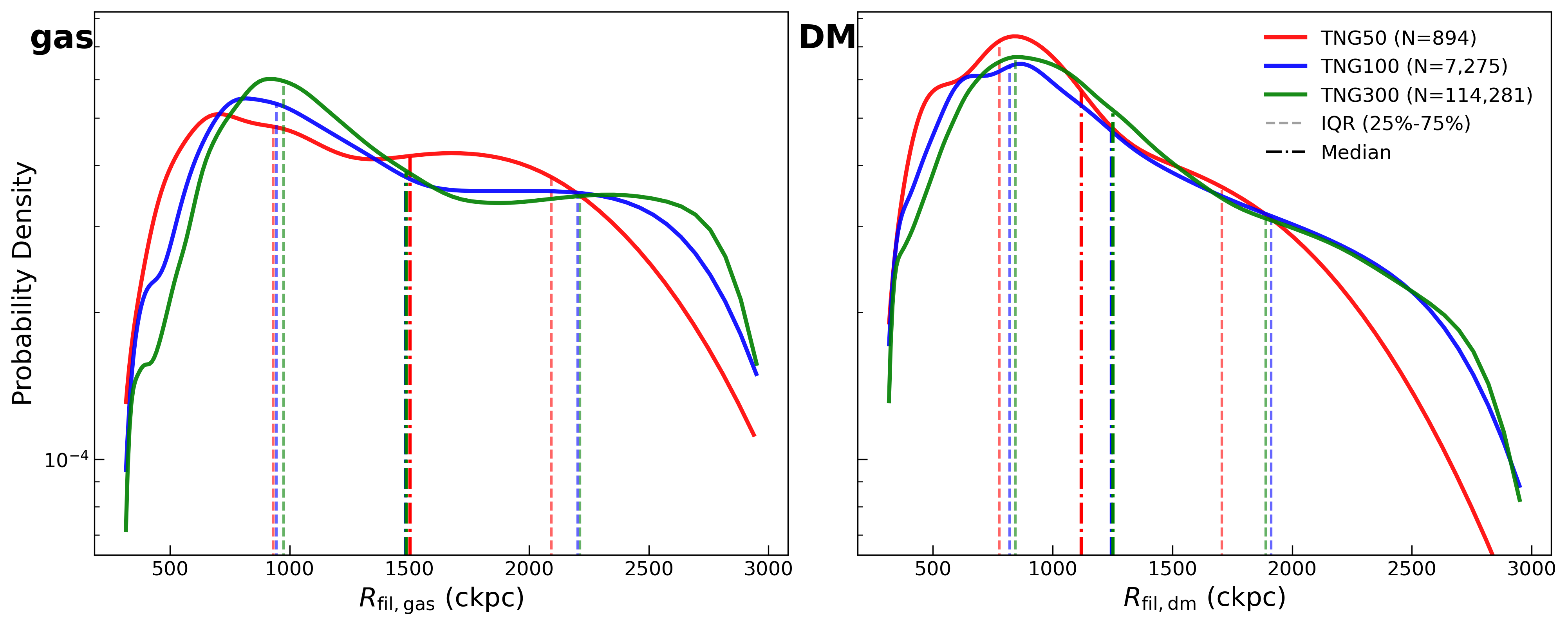}
\caption{The distribution of filament segment width in TNG50(red), TNG100(blue) and TNG300(green) at redshift $z=0$. Left and right panel shows results of width defined from baryonic and dark matter density profiles, respectively. The vertical dash-dotted lines mark the median, while the dashed lines mark 25th and 75th percentiles.}
\label{fig:width_pdf}
\end{centering}
\end{figure*}

To construct the temperature profile, the temperature of each gas cell in the IllustrisTNG simulation is derived from its InternalEnergy $u$ and ElectronAbundance $x_e$ using the following relation:
\begin{equation}
    T = (\gamma - 1)*\frac{{u}}{{k_B}}*\mu, 
\end{equation}
where $\gamma = 5/3$ is the adiabatic index, and, $k_B$ is the Boltzmann constant. The symbol $\mu$ denotes the mean molecular weight, which can be expressed as:
\begin{equation}
    \mu = \frac{\rm{4}}{{1 + 3\mathit{X}_H + 4\mathit{X}_H}x_e}*m_p, 
\end{equation}
where $\mathit{X}_H = 0.76$ is the hydrogen mass fraction, and $m_p$ is the proton mass. To derive the temperature profile of filament segments, we follow a procedure similar to that used for calculating the matter density profile—by computing the mass-weighted average temperature within each cylindrical shell.

The profiles of individual segments exhibit significant fluctuations. Therefore, in Section 4, we will present the mean density and temperature profiles by stacking filament segments with similar widths. 

\section{filament segment width-linear halo/stellar mass density relation} \label{sec:fila_width}

\subsection{width of filament segments at redshift 0}

Figure \ref{fig:width_pdf} presents the probability density function (PDF) of filament segment widths, measured from density profiles that exclude matter within halos massive than $10^{10}\, M_{\odot}$ at redshift $z=0$. If the width is instead derived from profiles that include halo matter, the values would be approximately $1.1-1.3$ times of those shown in Figure \ref{fig:width_pdf}. The left and right panels display the widths defined using baryonic and dark matter density profiles, respectively. The vertical dash-dotted and dashed lines indicate the median, 25th and 75th percentiles.

Most filament segments have widths between $0.4\,\mathrm{Mpc}$ and $2.75\,\mathrm{Mpc}$, with a plateau between $0.5$ and $1.2\,\mathrm{Mpc}$. Our analysis focuses on filaments within the approximate range of $0.3$ to $3.0\,\mathrm{Mpc}$, although segments narrower than $0.4\,\mathrm{Mpc}$ are likely underrepresented. The median width of our segment sample is approximately $1.2\,\mathrm{Mpc}$ and $1.5\,\mathrm{Mpc}$ for dark matter and gas component, respectively. These values are slightly larger than the value reported by \citet{2025arXiv250206484B}, who adopted a lower density threshold and included halo matter in their profile measurements. This discrepancy likely reflects differences in filament identification methods and selection criteria. The abundance of segments wider than $\sim 0.75\,\mathrm{Mpc}$ decreases steadily with increasing width, consistent with previous studies \citep{2014MNRAS.441.2923C, 2021ApJ...920....2Z, 2025arXiv250206484B}. This decline is more pronounced when widths are defined using dark matter ($R_{\mathrm{fil,dm}}$) compared to baryonic matter ($R_{\mathrm{fil,gas}}$), likely due to the smoothing effect of gas thermal pressure. Additionally, the smaller volume of TNG50 results in fewer thick filaments relative to TNG100 and TNG300, although its higher resolution enables the detection of narrower structures.

\begin{figure}[htb]
\begin{centering}
\hspace{-0.5cm}
\includegraphics[width=0.5\textwidth, clip]{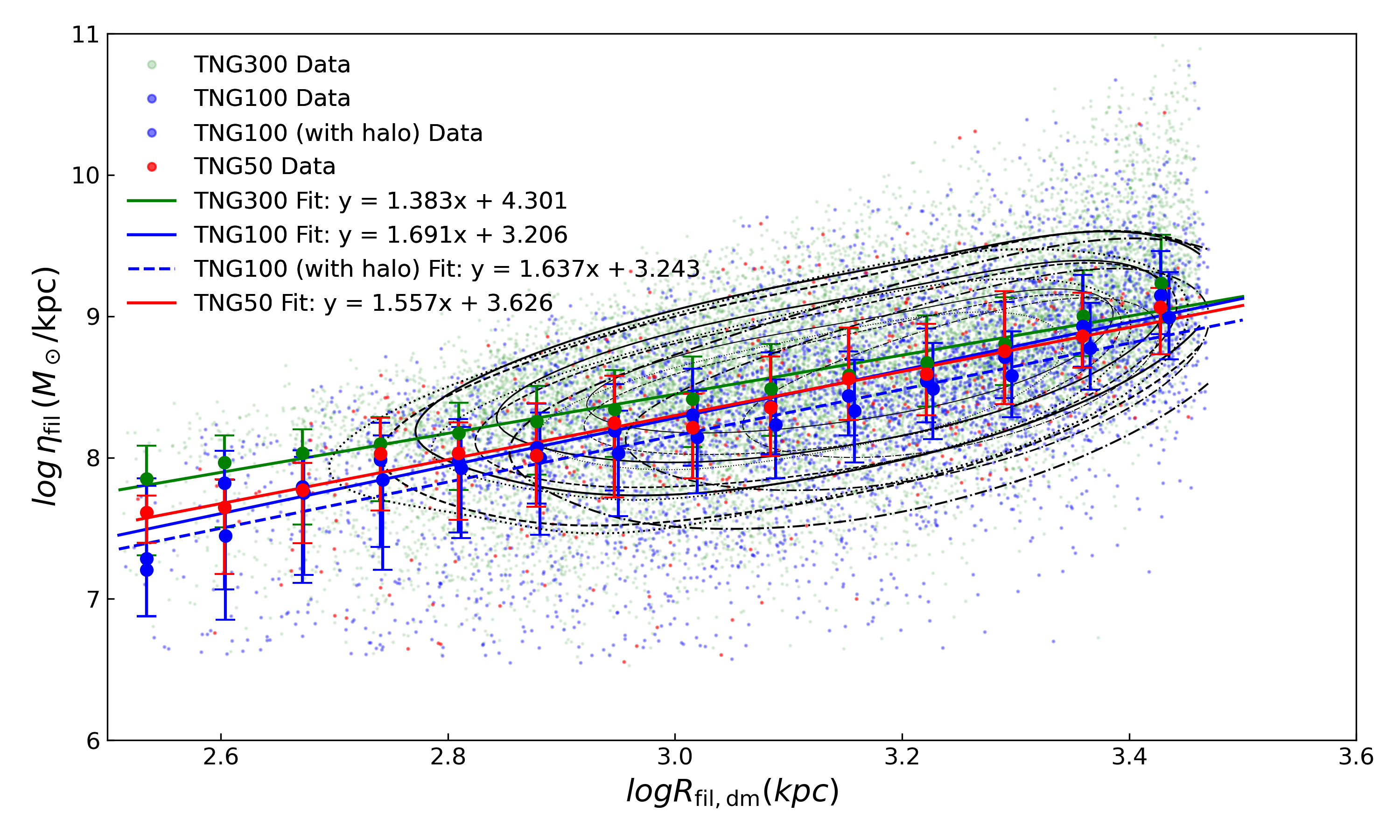}
\includegraphics[width=0.5\textwidth, clip]{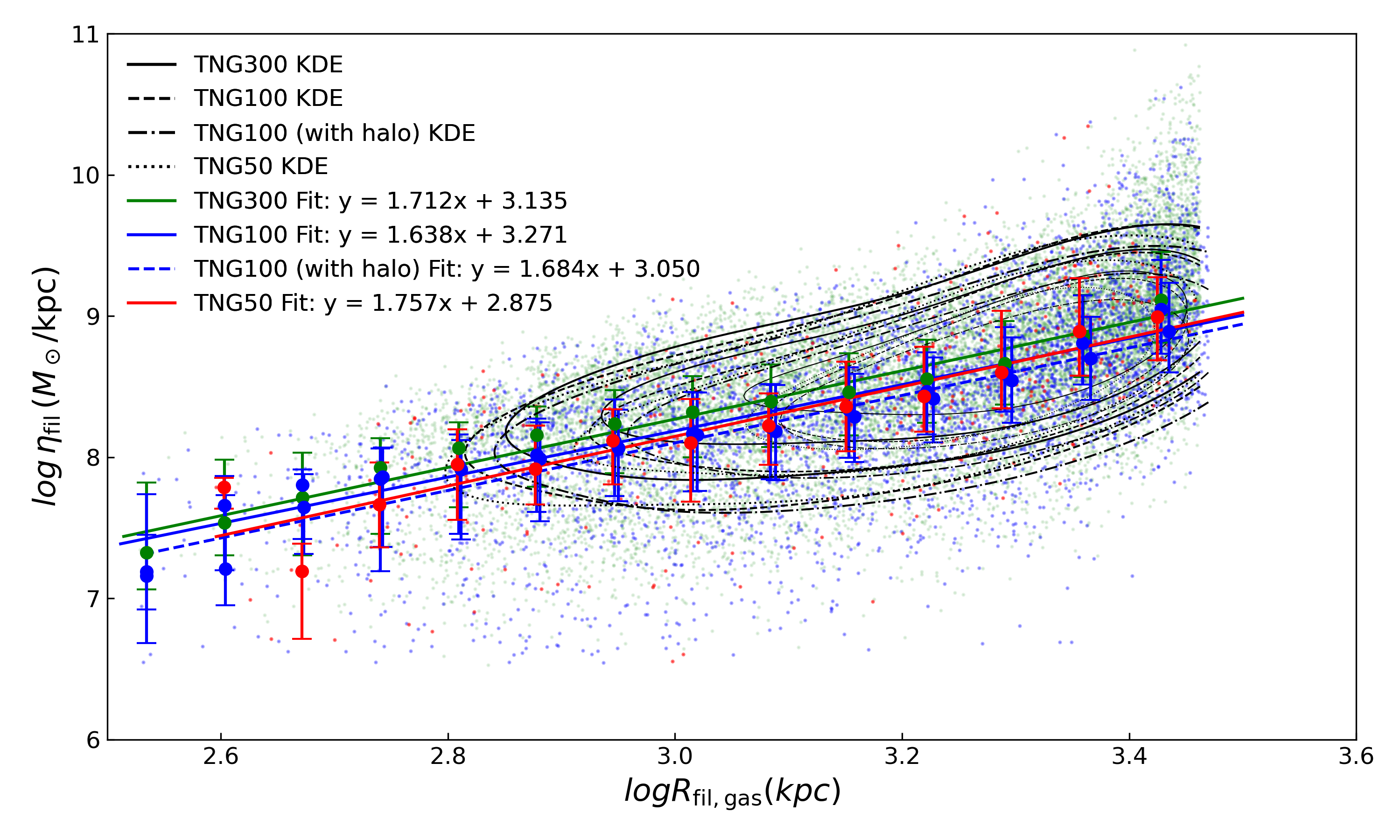}
\caption{Relationship between filament segment width and linear halo mass density across the TNG300 (green), TNG100 (blue), and TNG50 (red) simulations. The upper and lower panels show results using widths derived from dark matter and gas, respectively. Black solid, dashed, and dotted contours indicate the density distributions for each simulation. Lines with error bars represent the median relations with the interquartile range ($25\%$–$75\%$) for each simulation. The blue dashed line shows the fitted relation in TNG100 when the extended width—calculated by including halo matter in the density profile—is used.}
\label{fig:halomass_r}
\end{centering}
\end{figure}

\subsection{filament segment width - linear halo mass density relation}

\citet{2024ApJ...967..141Z} demonstrated a correlation between the width of cosmic filament segments and the linear halo mass density. Here, we further examine and confirm this relation using filament samples from the IllustrisTNG simulations. Following the approach of \citet{2024ApJ...967..141Z}, we define the linear halo mass density, $\rm{\eta_{fil}}$, as the total mass of halos embedded within each filament segment, divided by the segment’s length. We include halos with masses greater than $10^{10}\, \text{M}_{\odot}$ in this calculation. Figure~\ref{fig:halomass_r} shows the $\rm{\eta_{fil}}$ as a function of filament width for the TNG50, TNG100, and TNG300 simulations. Each dot represents an individual filament segment, colored according to its simulation. The top and bottom panels display the widths defined by the dark matter and baryonic matter density profiles, respectively.

Across all three simulations, we find an approximately linear correlation between filament width and the linear halo mass density, $\rm{\eta_{fil}}$. At the high-width end, this relation exhibits an upturn, typically associated with filament segments containing at least one massive halo ($M_{\mathrm{halo}}>10^{13}\, \mathrm{M}_{\odot}$). These trends are broadly consistent with the results of \citet{2024ApJ...967..141Z}, who reported a near-linear relationship within the width range of $1.5$–$4.5\,\mathrm{Mpc}$, with deviations at both the low and high $\rm{\eta_{fil}}$ ends. The overlaid contours in Figure~\ref{fig:halomass_r} represent the probability distribution of the data and show that widths defined using baryonic matter are systematically larger than those defined using dark matter. This systematic offset is likely due to the smoothing effect of gas thermal pressure, which broadens the baryonic matter distribution relative to the underlying dark matter.

We find that the median local linear halo mass density in each bin of filament width can be approximated as a function of the local filament width as follows:

\begin{equation}
    log_{10}(\frac{\rm{\eta_{fil}}}{\rm{\text{M}_{\odot}/kpc}}) \approx f_h*log_{10}\frac{\rm{R_{fil}}}{\rm{kpc}} + \beta_1
\end{equation}

where $f_h \sim 1.40$–$1.70$ and $\beta_1 \approx 3.6$ for $R_{fil,dm}$, and $f_h \sim 1.60$–$1.75$ and $\beta_1 \approx 3.0$ for $R_{fil,gas}$. As shown in Figure~\ref{fig:halomass_r}, the relationship between linear halo mass density and filament width holds consistently across the Illustris-TNG50, TNG100, and TNG300 simulations. The best-fitting coefficients are very similar among the three simulations and are indicated in the legend of Figure~\ref{fig:halomass_r}. In addition, solid straight lines show the fits for widths measured excluding halo matter, while the dashed straight line correspond to width measured including halo matter in TNG100. Results for TNG50 and TNG300, when the widths are measured including halo matter, show similar trends to TNG100, with only minor differences in magnitude and slope. To maintain clarity, their lines are not shown in this Figure. Overall, the correlation between $\eta_{fil}$ and $R_{fil}$ appears robust under both definitions of filament width and is only mildly sensitive to the simulation volume and resolution.

\subsection{filament segment width - linear stellar mass density relation}

\begin{figure}[htb]
\begin{centering}
\hspace{-0.5cm}
\includegraphics[width=0.5\textwidth, clip]{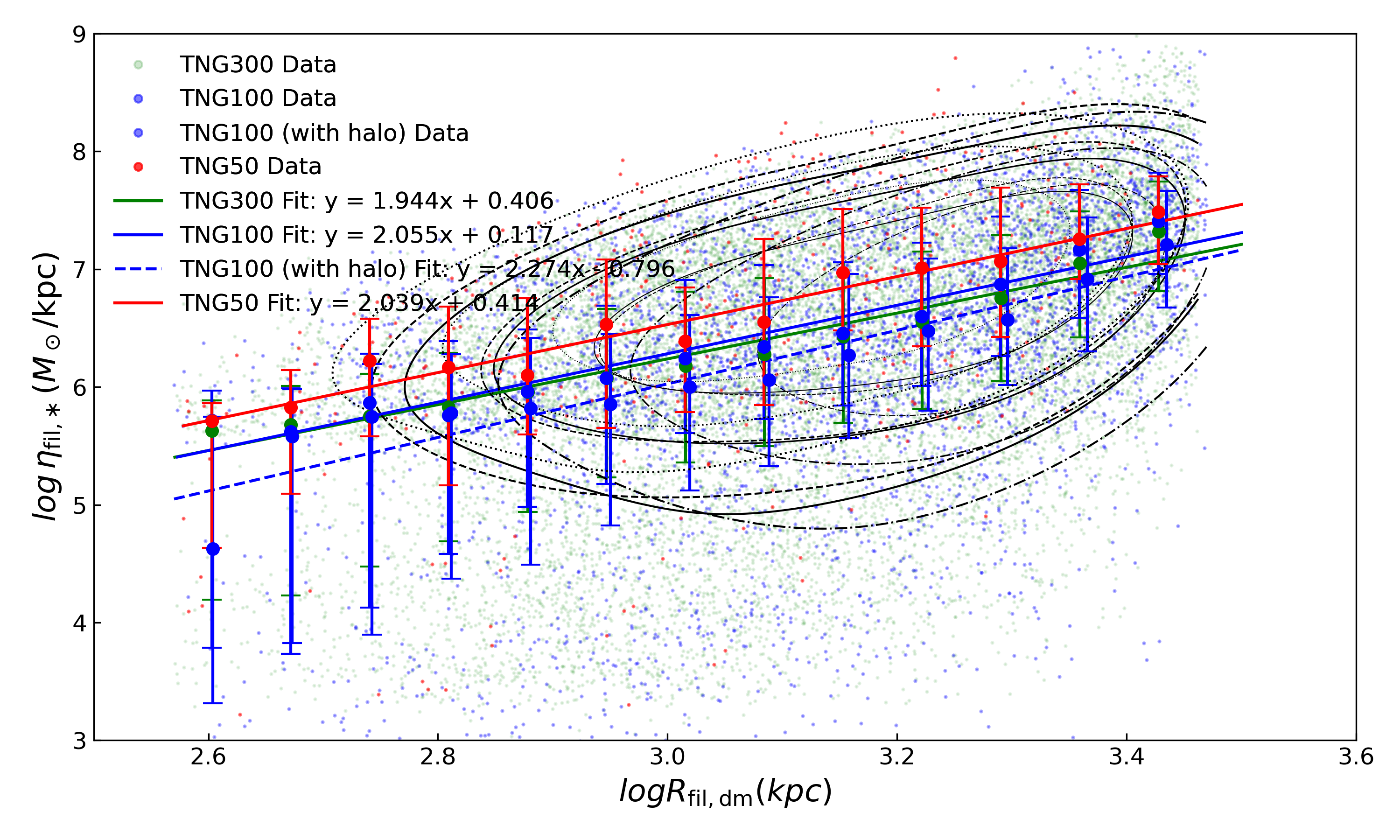}
\includegraphics[width=0.5\textwidth, clip]{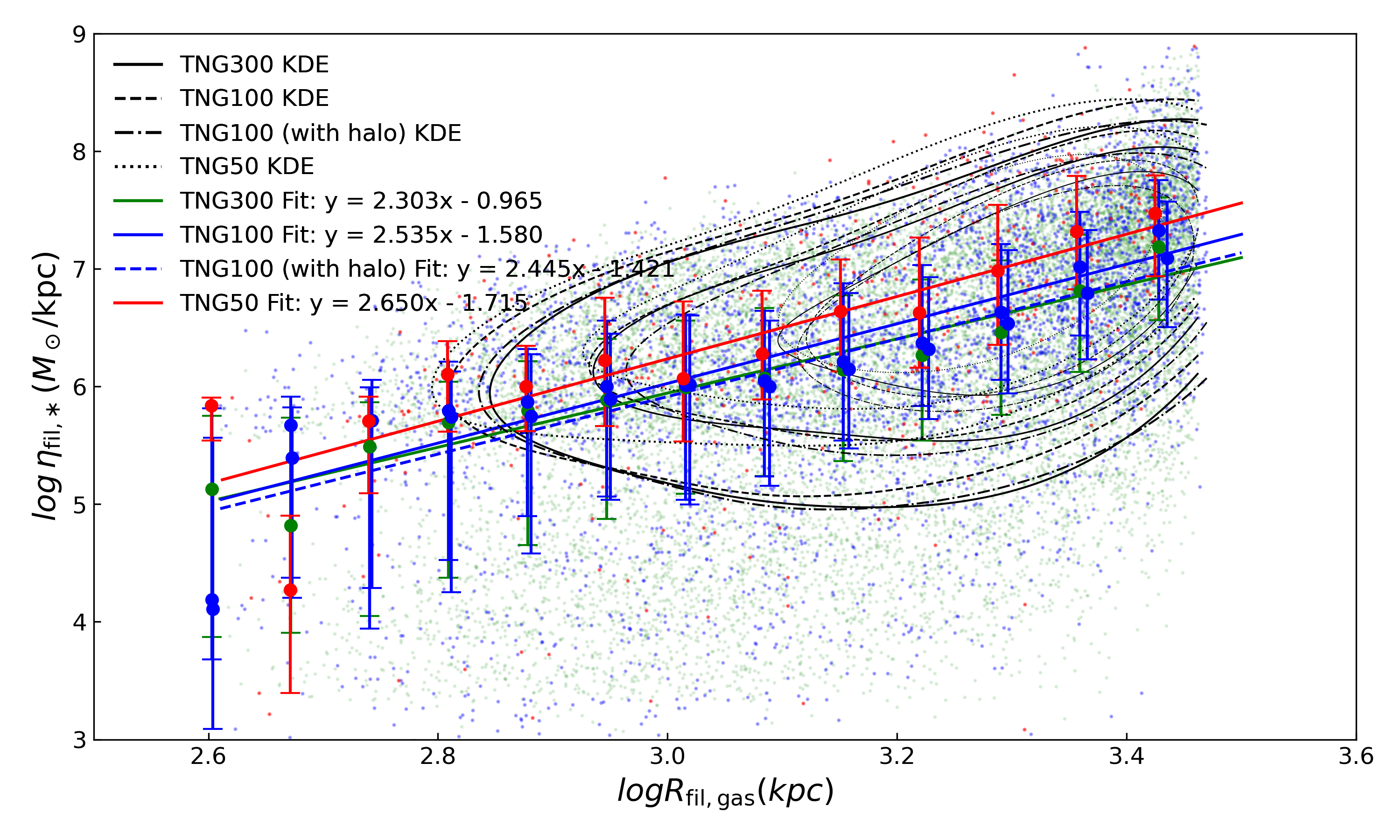}
\caption{Relation between filament segment width and linear stellar mass density across all three simulations. The upper and lower panels show results based on widths derived from dark matter and gas, respectively. Colored points and error bars represent data from TNG300 (green), TNG100 (blue), and TNG50 (red). The contours indicate the density distribution of segments, with different line styles corresponding to different simulations.}
\label{fig:stellarmass_r}
\end{centering}
\end{figure}

The correlation between linear halo mass density and filament segment width cannot be directly applied to observational data, as it is challenging to measure the halo mass of individual galaxies. However, a correlation exists between stellar mass and halo mass in the universe (e.g., \citealt{2020A&A...634A.135G}). Given that the IllustrisTNG simulations include stellar components, it is possible to investigate the relationship between the stellar mass per unit length within filament segments and their width.

To this end, we calculate the linear stellar mass density, $\eta_{\mathrm{fil},\ast}$, for each filament segment using a method analogous to that used for the linear halo mass density, but considering all the stellar mass in halos that with total mass greater than $10^{10}\, \mathrm{M}_{\odot}$ and located within the segments. As shown in Figure \ref{fig:stellarmass_r}, $\eta_{\mathrm{fil},\ast}$ also exhibits a roughly linear correlation with filament width. However, the scatter in the $\eta_{\mathrm{fil},\ast}$–$R_{\mathrm{fil}}$ relation is more pronounced than in the $\eta_{\mathrm{fil}}$–$R_{\mathrm{fil}}$ case, likely due to the intrinsic scatter in the stellar mass–halo mass relation.

We also perform linear fits of the linear stellar mass density, $\eta_{\mathrm{fil},\ast}$ as a function of filament width, defined with/without matters in halos, and find that the slopes are relatively consistent across different simulations. TNG50 and TNG300 results, including halo matter, show similar trends but are omitted from the figure for clarity. However, the intercepts show moderate variation. The median value of $\eta_{\mathrm{fil},\ast}$ in each width bin can be approximately described by the following relation:

\begin{equation}
    log_{10}(\frac{\rm{\eta_{fil,\ast}}}{\rm{\text{M}_{\odot}/kpc}}) \approx f_s*log_{10}\frac{\rm{R_{fil}}}{\rm{kpc}} + \beta_2 ,
\label{eqn:stellar_mass_r}
\end{equation}

where $f_s \sim 1.95$–$2.05$ for $R_{\mathrm{fil},\mathrm{dm}}$ and $f_s \sim 2.3$–$2.6$ for $R_{\mathrm{fil},\mathrm{gas}}$. While the values of $f_s$ are fairly consistent across the three IllustrisTNG simulations, the parameter $\beta_2$ exhibits moderate variation, with TNG50 showing the highest value. This trend is likely due to the increased star formation efficiency observed at higher resolutions within the IllustrisTNG project \citep{2018MNRAS.473.4077P}, despite the sub-grid models for star formation and feedback remaining largely unchanged across simulations \citep{2019MNRAS.490.3234N}.

The relation presented in Equation \ref{eqn:stellar_mass_r} offers a practical tool for estimating the local width of filaments, that is width of filament segments, identified from the observed galaxy distribution. This is particularly valuable for probing the missing baryons in the cosmic web and investigating how the large-scale filamentary structure influences galaxy properties.

\begin{figure*}[htb]
\begin{centering}
\hspace{-1.0cm}
\includegraphics[width=1.1\textwidth, clip]{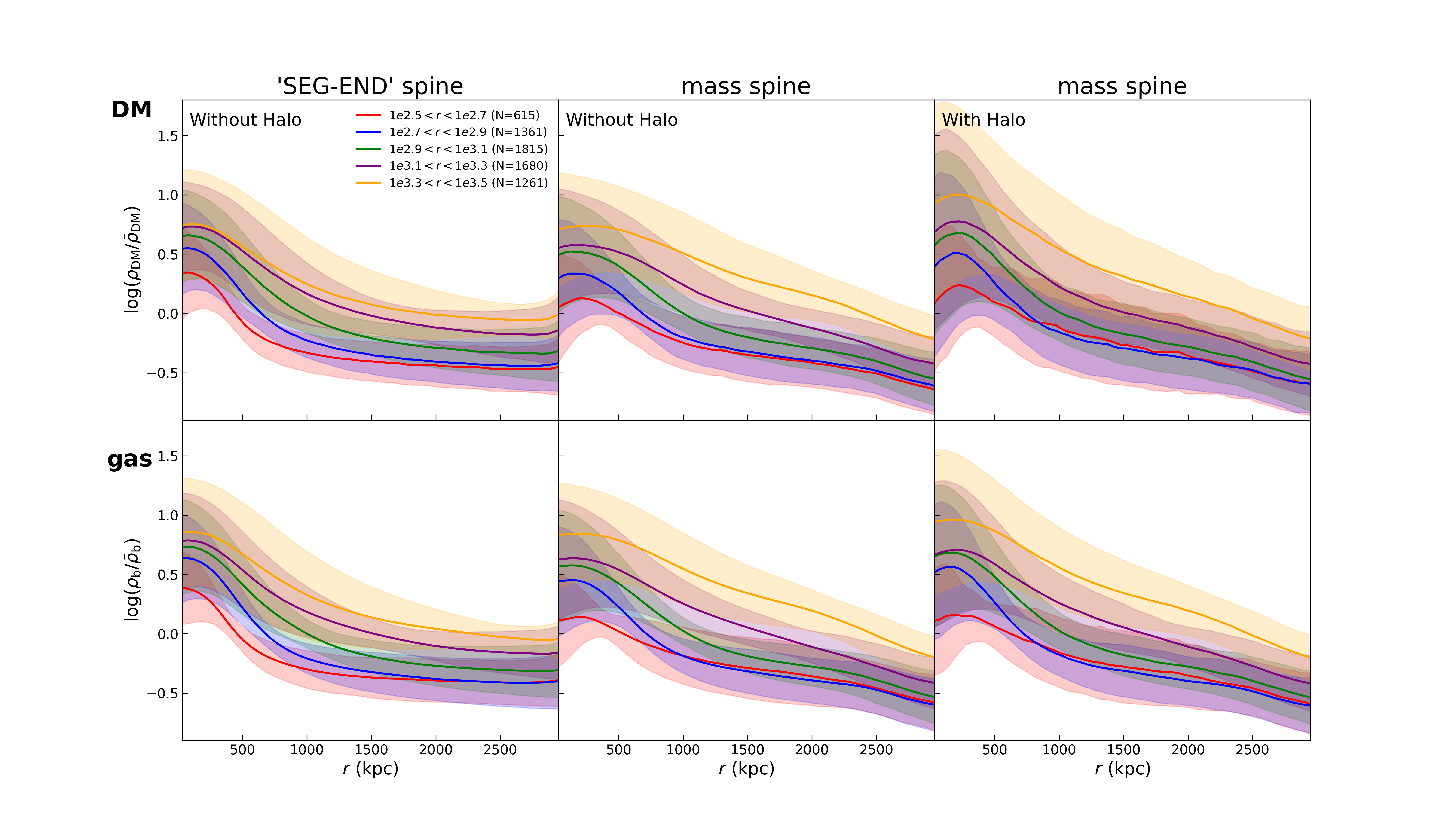}
\caption{Solid lines represent the mean density profiles of filament segments in five width bins from the TNG100 simulation at redshift $z = 0$. Red, blud, green, purple and yellow color indicate width of $10^{2.5}-10^{2.7}$, $10^{2.7}-10^{2.9}$, $10^{2.9}-10^{3.1}$, $10^{3.1}-10^{3.3}$ and $10^{3.3}-10^{3.5}$ kpc, respectively. Shaded regions represent the standard deviations. The upper panels show the mean dark matter density profiles, while the lower panels show the mean baryon density profiles. The left and middle panels correspond to filaments excluding massive halos ($M_{\mathrm{halo}} > 10^{10},\mathrm{M_\odot}$), using the SEG-END spine and mass spine as the axis, respectively. The right panels are based on the mass spine and include contributions from halo matter. }
\label{fig:denprof_tng100}
\end{centering}
\end{figure*}

\begin{figure*}[htbp]
\begin{centering}
\hspace{-1.0cm}
 \includegraphics[width=1.1\textwidth, clip]{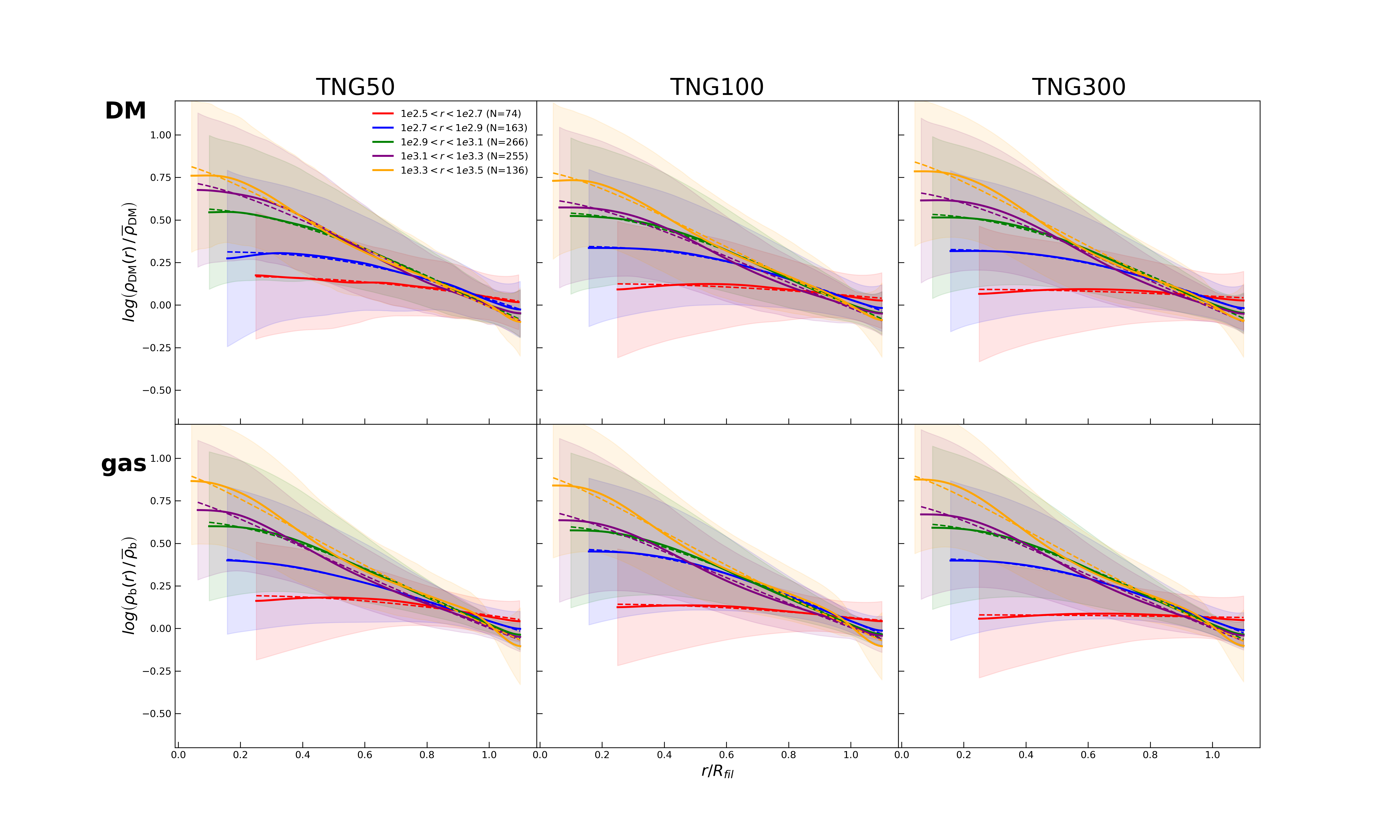}    
\caption{Rescaled density profiles across the three IllustrisTNG simulations and different filament width bins. The top three panels display dark matter density profiles, where solid lines represent the mean profiles, shaded regions indicate the standard deviation, and dashed lines correspond to the best-fit model Equation~\ref{eqn:den_fit_all}. The bottom three panels show the gas density profiles, following the same labeling conventions as the upper panels.}
    \label{fig:rdm_rgas_combined}
\end{centering}
\end{figure*}

\section{Density and temperature profiles}

\subsection{density profiles}
Filaments with different widths exhibit distinct density profiles, characterized by the variation of density as a function of absolute distance from the filament spine (e.g., \citealt{2014MNRAS.441.2923C, 2021ApJ...920....2Z, 2025arXiv250206484B}). However, \citet{2021ApJ...920....2Z} found that when these profiles are normalized by the filament width, they display self-similar behavior and can be well described by a single isothermal $\beta$-model. In this work, we test this feature using filament segments from the IllustrisTNG simulations. We divide the segments into five groups based on their local widths, spanning the range from $10^{2.5}$ to $10^{3.5},\mathrm{kpc}$, and compute the mean density profile for each group.

Figure \ref{fig:denprof_tng100} presents the resulting profiles for filament segments in the TNG100 simulation. The left and middle columns show density profiles excluding halo matter, using the stellar and mass spines as the axis, respectively, while the right column includes halo matter and uses the mass spine. Top and bottom rows display dark matter and baryonic matter mean profiles. Thicker filaments exhibit higher central densities and more extended profiles. When halo matter is excluded, the central density peaks at $\sim 5-15$ in the thickest group and decreases to $\sim 1-3$ in the thinnest. Using the mass spine yields broader profiles, and including halo matter increases central densities by $\sim 30\%$. In addition, thicker filaments show shallower radial declines, reaching the cosmic mean at larger distances. This behavior likely arises because thick filaments tend to reside in regions of higher large-scale overdensity. As a result, their density transitions more gradually toward the cosmic mean and underdense (void) regions, compared to thinner filaments. Despite these trends, individual filament segments exhibit significant fluctuations, and noticeable variations remain among segments within the same width group. These results are generally in agreement with \citet{2014MNRAS.441.2923C}, \citet{2021ApJ...920....2Z} and \citet{2025arXiv250206484B}

We further examine density profiles as a function of radius normalized by the segment width, denoted as `rescaled density profiles', focusing on profiles aligned with the mass spine and excluding halo contributions. As shown in Figure~\ref{fig:rdm_rgas_combined}, segments with widths greater than $0.8\,\mathrm{Mpc}$ exhibit strong self-similarity in their rescaled density profiles, particularly in the outer regions ($\sim0.4$–$1.2\,R_{\mathrm{fil}}$). Some deviations appear in the inner cores, especially for relatively thinner segments.  These deviations are likely due in part to the limited resolution of the simulations, which makes it challenging to accurately capture the matter distribution in the inner regions of thin filaments. Additional contributing factors include small sample sizes and the fact that each width bin spans a range of 0.2 dex. Furthermore, physical processes such as baryonic feedback may also play a role in producing these deviations. These discrepancies are less pronounced in TNG100, especially for the two most populated groups. The observed self-similarity aligns with the findings of \citet{2021ApJ...920....2Z}, who reported consistent behavior for filaments with widths between $1$–$4\,\mathrm{Mpc}/h$. In addition, Figure~\ref{fig:rdm_rgas_combined} shows that the rescaled density profiles are broadly consistent across TNG50, TNG100, and TNG300, for both dark matter and baryonic components.

Moreover, the average density profiles across all five width-based groups can be reasonably well described by a form as 
\begin{equation}
    \rho_{fil} (r) = \frac{{\rho_{0}}}{{(1 + (\frac{\rm{r{}}}{{R_{fil}}})^\alpha)^{\frac{3}{2}\beta}}}
\label{eqn:den_fit_all}
\end{equation}
Here, $\rho_{\mathrm{fil}}(r)$ represents the density of dark matter or baryonic matter. The parameter $\rho_{0}$ corresponds to the central (core) density of the filament, and $R_{\mathrm{fil}}$ denotes the filament segment width. When $\alpha = 2$, the expression reduces to the standard isothermal $\beta$-model. The best-fit values of $\rho_{0}$, $\alpha$, and $\beta$ for each width group are listed in Table~\ref{tab:fit_quality}. Notably, the fitted central density $\rho_{0}$ closely matches the average density measured in the innermost radial bin.

Nevertheless, the fitted values of $\alpha$ and $\beta$ show moderate deviations from those reported by \citet{2021ApJ...920....2Z}, where $\alpha=2$ and $\beta=2/3$. Specifically, $\alpha$ tends to decrease with increasing $R_{\mathrm{fil}}$, while $\beta$ shows the opposite trend in this work. These differences are likely attributable to variations in gravity and hydrodynamical solvers, and baryonic physics between the simulations—AREPO in IllustrisTNG versus RAMSES in \citet{2021ApJ...920....2Z}—as well as differences in filament identification methods and profile construction procedures. For filament segments with widths between $0.8$ and $2.0,\mathrm{Mpc}$, which account for approximately $50\%-60\%$ of our sample, $\alpha$ remains close to 2, consistent with the isothermal $\beta$-model and the fitting form used by \citet{2021ApJ...920....2Z}. However, the corresponding values of $\beta$ in our analysis are higher.

The universal fitting function given in Equation~\ref{eqn:den_fit_all} effectively captures the density profiles across all filament width groups and simulations, as shown in Figure~\ref{fig:rdm_rgas_combined}. In the figure, solid lines indicate the average density profiles, while dashed lines represent the corresponding best-fit curves from Equation~\ref{eqn:den_fit_all}. The goodness of fit, measured by $\chi^2$ values typically ranging from 0.1 to 1.0, indicates generally good agreement. We also find that, when halo matter is included, density profiles computed using the mass spine as the axis remain well described by the same fitting function, consistent with \citet{2021ApJ...920....2Z}. The corresponding fit parameters for TNG100 are provided in Table~\ref{tab:fit_quality}. In contrast, profiles computed with the SEG-END spine as the axis are poorly described by this model.

\begin{table}[htb]
\hspace{-1.0cm}
\centering
\hspace{-1.0cm}
\caption{Fitting parameters of rescaled filament density profiles described by Equation~\ref{eqn:den_fit_all}. The terms`dm' and `gas' indicate profiles of dark matter and gas respectively. Matters in halos are excluded, except for the last two sets (`with halo'). The degrees of freedom (dof) vary by width group and are 4, 9, 17, 30, and 48, calculated as the number of radial bins minus three. Thinner groups have fewer bins and thus lower dof.} 
\label{tab:fit_quality}
\renewcommand{\arraystretch}{0.75}
\setlength{\tabcolsep}{0.6em}
\begin{tabular}{l c c c c c c}
\toprule
\multirow{2}{*}{Samples} & \multicolumn{5}{c}{fitting parameters} \\
\cmidrule(lr){2-7}
 & $ log(R_{\rm fil})$  &$\alpha$ & $\beta$ & $\rho_{\rm 0,fit}$ & $\rho_{\rm 0}$ &$\chi^2$ \\
\midrule
\multirow{6}*{$\rm{TNG50(dm)}$} 
 & $2.5-2.7$ & 3.00 & 0.27 & 1.48 & 1.66 & 0.10 \\
 & $2.7-2.9$ & 3.00 & 0.61 & 2.06 & 2.11 & 0.17 \\
 & $2.9-3.1$ & 2.10 & 1.25 & 3.72 & 4.11 & 0.37 \\
 & $3.1-3.3$ & 1.55 & 1.64 & 5.34 & 5.29 & 0.95 \\
 & $3.3-3.5$ & 1.33 & 1.87 & 6.77 & 7.24 & 0.57 \\
 \cmidrule{2-7}
 \multirow{6}*{$\rm{TNG50(gas)}$} 
 & $2.5-2.7$ & 3.00 & 0.26 & 1.57 & 1.40 & 0.52 \\
 & $2.7-2.9$ & 2.26 & 0.81 & 2.59 & 2.61 & 0.05 \\
 & $2.9-3.1$ & 1.92 & 1.35 & 4.31 & 4.68 & 0.40 \\
 & $3.1-3.3$ & 1.30 & 1.70 & 5.90 & 5.67 & 0.70 \\
 & $3.3-3.5$ & 1.28 & 2.00 & 8.27 & 8.08 & 2.08 \\
\midrule
\multirow{6}*{$\rm{TNG100(dm)}$}
 & $2.5-2.7$ & 3.00 & 0.15 & 1.34 & 1.20 & 0.28 \\
 & $2.7-2.9$ & 2.93 & 0.67 & 2.22 & 2.44 & 0.08 \\
 & $2.9-3.1$ & 2.14 & 1.20 & 3.51 & 3.75 & 0.43 \\
 & $3.1-3.3$ & 1.61 & 1.42 & 4.19 & 4.32 & 1.10 \\
 & $3.3-3.5$ & 1.46 & 1.76 & 6.12 & 5.93 & 0.36 \\
 \cmidrule{2-7}
 \multirow{6}*{$\rm{TNG100(gas)}$}
 & $2.5-2.7$ & 3.00 & 0.17 & 1.39 & 1.29 & 0.12 \\
 & $2.7-2.9$ & 2.53 & 0.94 & 2.94 & 3.14 & 0.20 \\
 & $2.9-3.1$ & 1.97 & 1.29 & 4.03 & 4.19 & 0.55 \\
 & $3.1-3.3$ & 1.39 & 1.53 & 4.97 & 4.92 & 0.58 \\
 & $3.3-3.5$ & 1.31 & 1.98 & 8.05 & 7.62 & 1.85 \\
\midrule
\multirow{6}*{$\rm{TNG300(dm)}$}
 & $2.5-2.7$ & 3.00 & 0.09 & 1.24 & 1.20 & 0.29 \\
 & $2.7-2.9$ & 3.00 & 0.64 & 2.13 & 2.43 & 0.03 \\
 & $2.9-3.1$ & 2.21 & 1.17 & 3.44 & 3.92 & 0.37 \\
 & $3.1-3.3$ & 1.61 & 1.53 & 4.68 & 4.96 & 1.62 \\
 & $3.3-3.5$ & 1.35 & 1.93 & 7.20 & 6.97 & 0.82 \\
 \cmidrule{2-7}
 \multirow{6}*{$\rm{TNG300(gas)}$}
 & $2.5-2.7$ & 3.00 & 0.03 & 1.20 & 1.19 & 0.29 \\
 & $2.7-2.9$ & 2.69 & 0.81 & 2.58 & 2.84 & 0.08 \\
 & $2.9-3.1$ & 1.99 & 1.33 & 4.17 & 4.54 & 0.60 \\
 & $3.1-3.3$ & 1.41 & 1.63 & 5.45 & 5.62 & 1.31 \\
 & $3.3-3.5$ & 1.34 & 2.00 & 8.20 & 8.62 & 1.83 \\
\midrule
\multirow{6}*{\shortstack[l]{$\rm{TNG100(dm)}$\\(with halo)}}
 & $2.5-2.7$ & 3.00 & 0.19 & 1.68 & 1.16 & 0.16 \\
 & $2.7-2.9$ & 2.75 & 0.99 & 3.29 & 2.59 & 0.22 \\
 & $2.9-3.1$ & 1.68 & 1.67 & 5.52 & 3.71 & 0.65 \\
 & $3.1-3.3$ & 1.21 & 2.00 & 7.24 & 4.70 & 1.07 \\
 & $3.3-3.5$ & 1.38 & 2.00 & 8.26 & 7.39 & 1.85 \\
 \cmidrule{2-7}
 \multirow{6}*{\shortstack[l]{$\rm{TNG100(gas)}$\\(with halo)}}
 & $2.5-2.7$ & 3.00 & 0.11 & 1.44 & 1.20 & 0.03 \\
 & $2.7-2.9$ & 2.32 & 1.21 & 3.95 & 3.42 & 0.27 \\
 & $2.9-3.1$ & 1.73 & 1.61 & 5.50 & 4.57 & 0.64 \\
 & $3.1-3.3$ & 1.25 & 1.77 & 6.28 & 5.11 & 0.59 \\
 & $3.3-3.5$ & 1.33 & 2.00 & 8.59 & 8.37 & 2.16 \\
\bottomrule
\end{tabular}
\end{table}

The success of Equation~\ref{eqn:den_fit_all} in capturing the density profiles of cosmic filaments underscores their self-similar nature. The consistency of these profiles across different IllustrisTNG simulations, and their agreement with results from simulations using other code (e.g., \citealt{2021ApJ...920....2Z}), further supports the robustness of this feature. This self-similarity likely arises from the gravitational collapse and partial virialization of matter along the radial direction of filaments, shaping their structure in a universal manner.

\begin{figure}[htbp]
\begin{centering}
\hspace{-0.5cm}
\includegraphics[width=0.5\textwidth, clip]{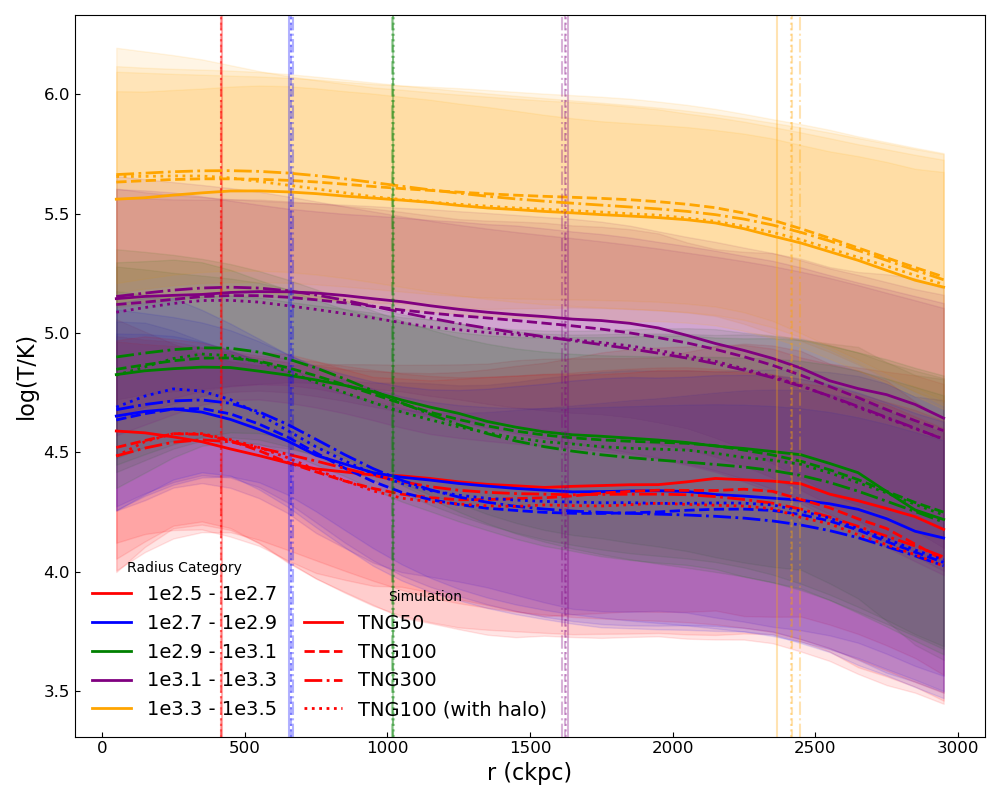}
\caption{Mass-weighted mean temperature profiles of filament segments grouped into five width bins ranging from $10^{2.5}$ to $10^{3.5}\,\mathrm{kpc}$. Solid, dashed, and dash-dotted lines represent results from the TNG50, TNG100, and TNG300 simulations, respectively. The dotted line shows the corresponding result from TNG100 when halo matter is included. Shaded regions indicate the standard deviation, and vertical lines mark the characteristic width of each width bin. }
\label{fig:T_profile}
\end{centering}
\end{figure}

\subsection{temperature profiles}
Figure \ref{fig:T_profile} shows the mass-weighted mean temperature profiles of filaments in the three simulations. The temperature profiles show notable fluctuations among individual filaments in the same width groups. We further calculate the mass weighted average temperature profiles for filaments segments in the five width groups. Generally, the average gas temperature drops first very slowly from the spine toward outside. The decline become more significant at a radius around $R_{fil,gas}$. Therefore, the regime within $R_{fil,gas}$ can be considered as nearly isothermal, similar to \cite{2021ApJ...920....2Z} for segments with similar width. A central isothermal core in filaments has also reported in \citet{2021A&A...649A.117G}, \citet{2021A&A...646A.156T}, \citet{2021MNRAS502351R}, \citet{2024MNRAS.52711256L} and \citet{2025arXiv250206484B}, but exhibits various radial ranges in these studies. The probable reasons include differences in simulations and classification of filaments.

The average temperature profiles from the TNG50, TNG100, and TNG300 simulations show good agreement. Including halo matter results in only mild changes. Thicker filaments tend to be hotter, consistent with trends reported by \citet{2021ApJ...920....2Z} and \citet{2025arXiv250206484B}. For widths between $10^{3.3}$ and $10^{3.5}\,\mathrm{kpc}$, the mean gas temperature exceeds $10^{5.5}\,\mathrm{K}$, making them promising candidates for detecting missing baryons. In contrast, filaments with widths between $10^{2.5}$ and $10^{2.7}\,\mathrm{kpc}$ have mean temperatures around $10^{4.5}\,\mathrm{K}$. These values are consistent with \citet{2021A&A...646A.156T} and \citet{2025arXiv250206484B}, but about $0.2$–$0.3$ dex lower than those in \citet{2021ApJ...920....2Z}.

\begin{figure*}[htb]
\begin{centering}
\hspace{-0.0cm}
\includegraphics[width=1.0\textwidth, clip]{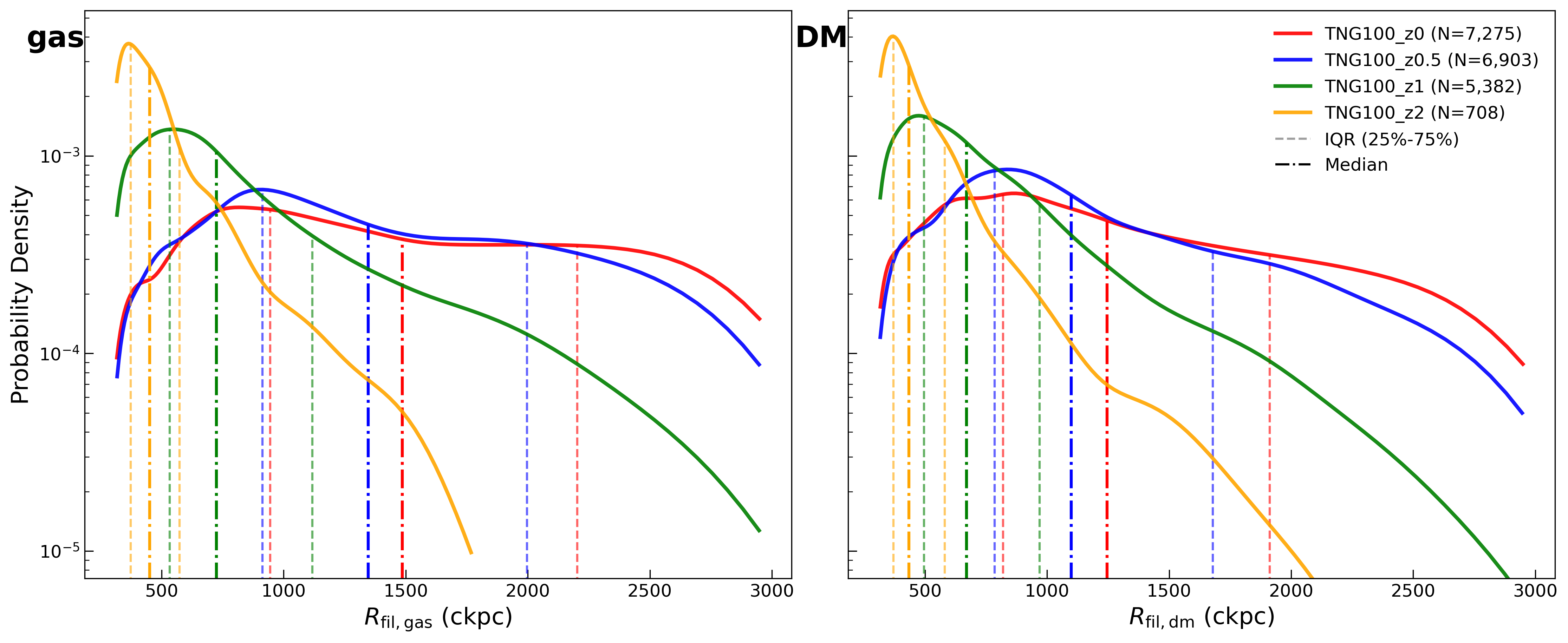}
\caption{The distribution of segment width in TNG100 at redshift $z=0$ (red), $z=0.5$ (blue), $z=1.0$ (green) and $z=2.0$ (orange). Left and right panel shows results of width defined from gas and dark matter density profiles, respectively. The vertical dash-dotted lines mark the median, while the dashed lines mark 25th and 75th percentiles.}
\label{fig:width_pdf_z}
\end{centering}
\end{figure*}

\begin{figure*}[htb]
\begin{centering}
\hspace{-0.0cm}
\includegraphics[width=0.45\textwidth, clip]{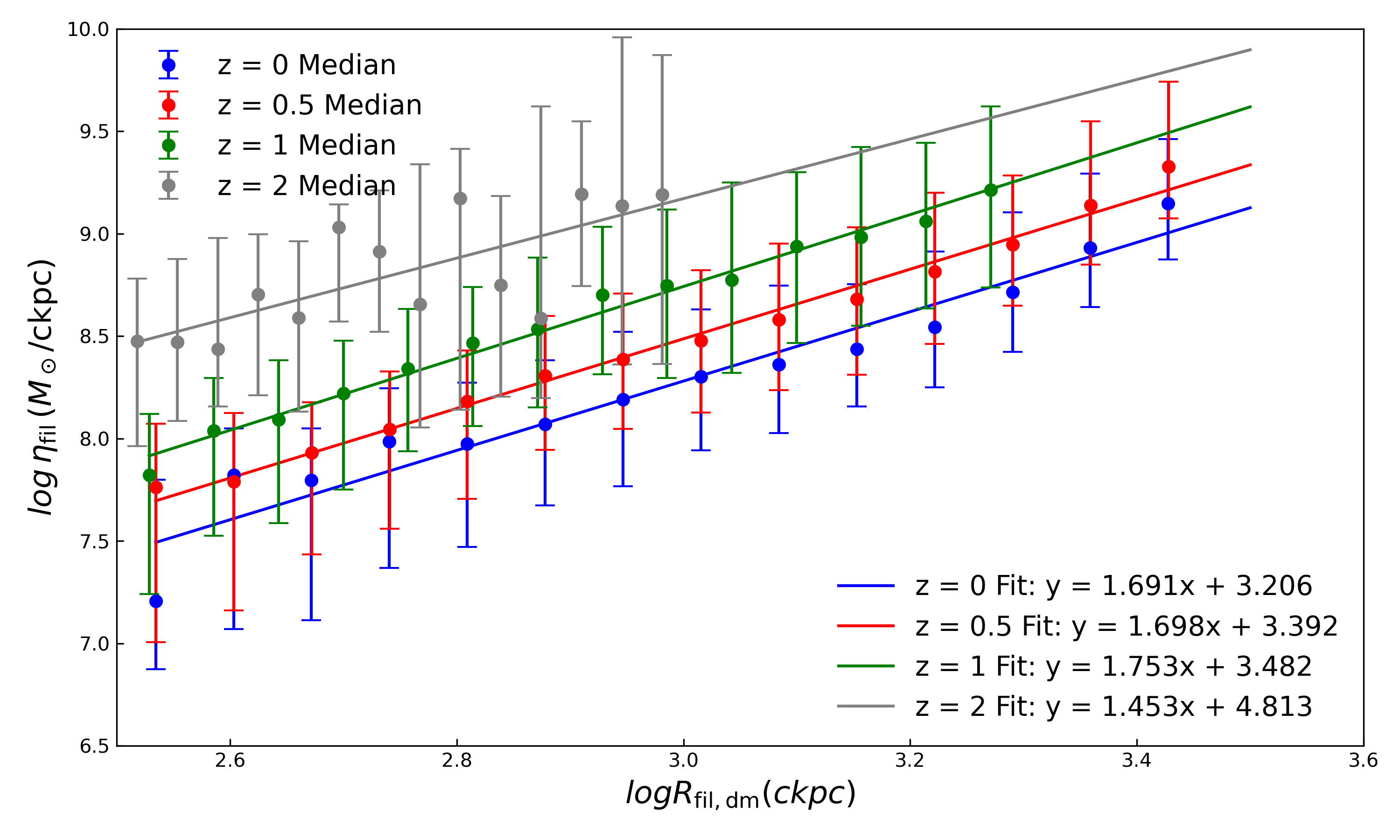}
\includegraphics[width=0.45\textwidth, clip]{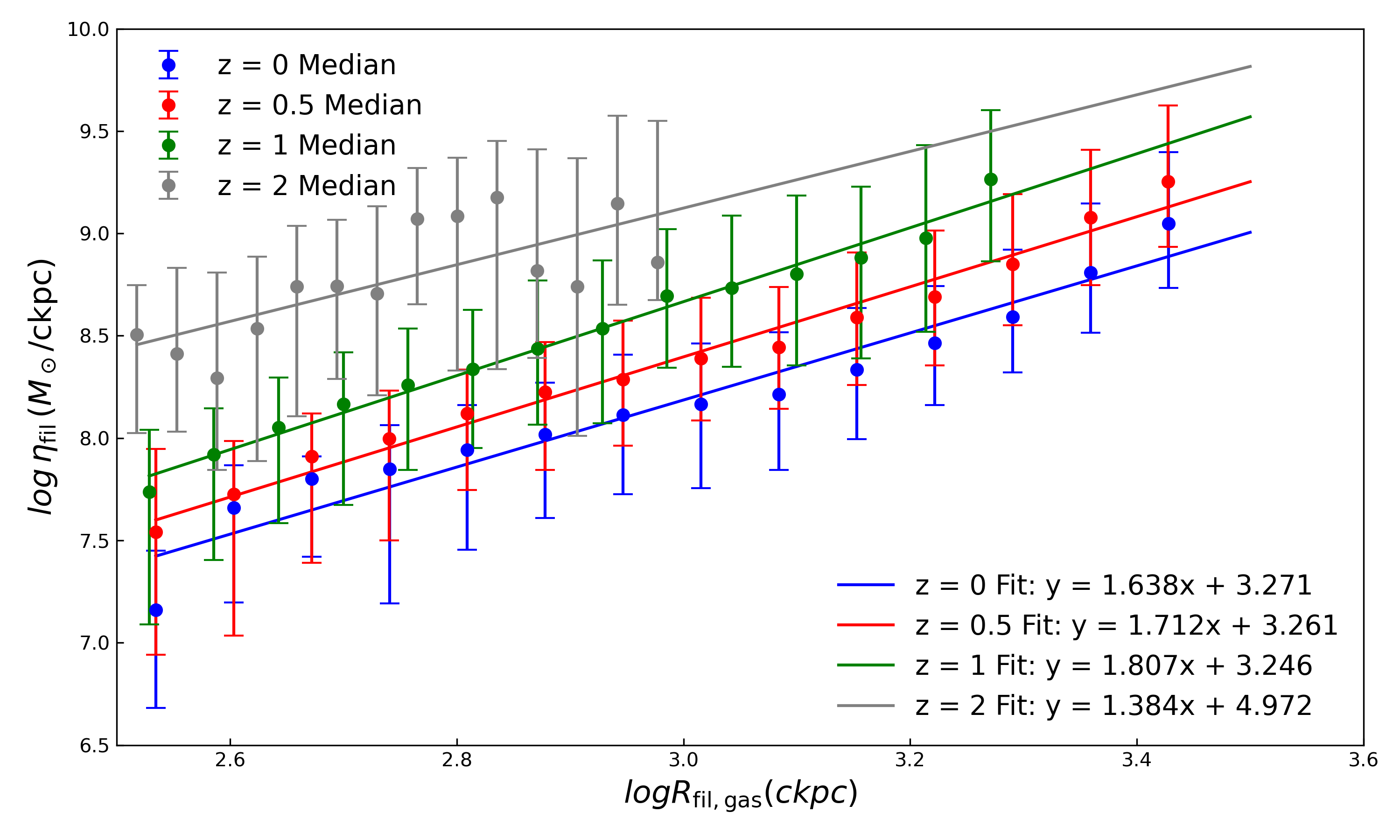}
\includegraphics[width=0.45\textwidth, clip]{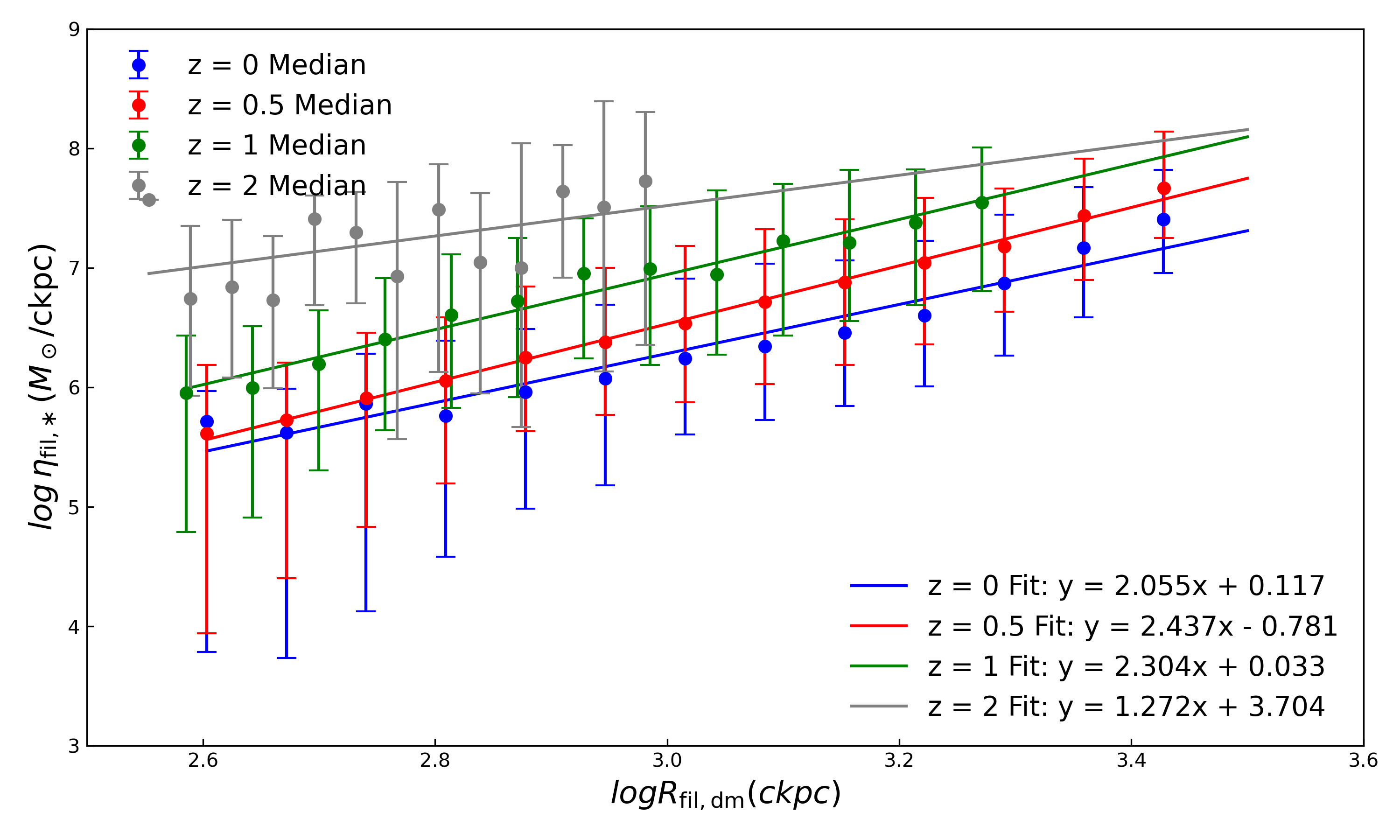}
\includegraphics[width=0.45\textwidth, clip]{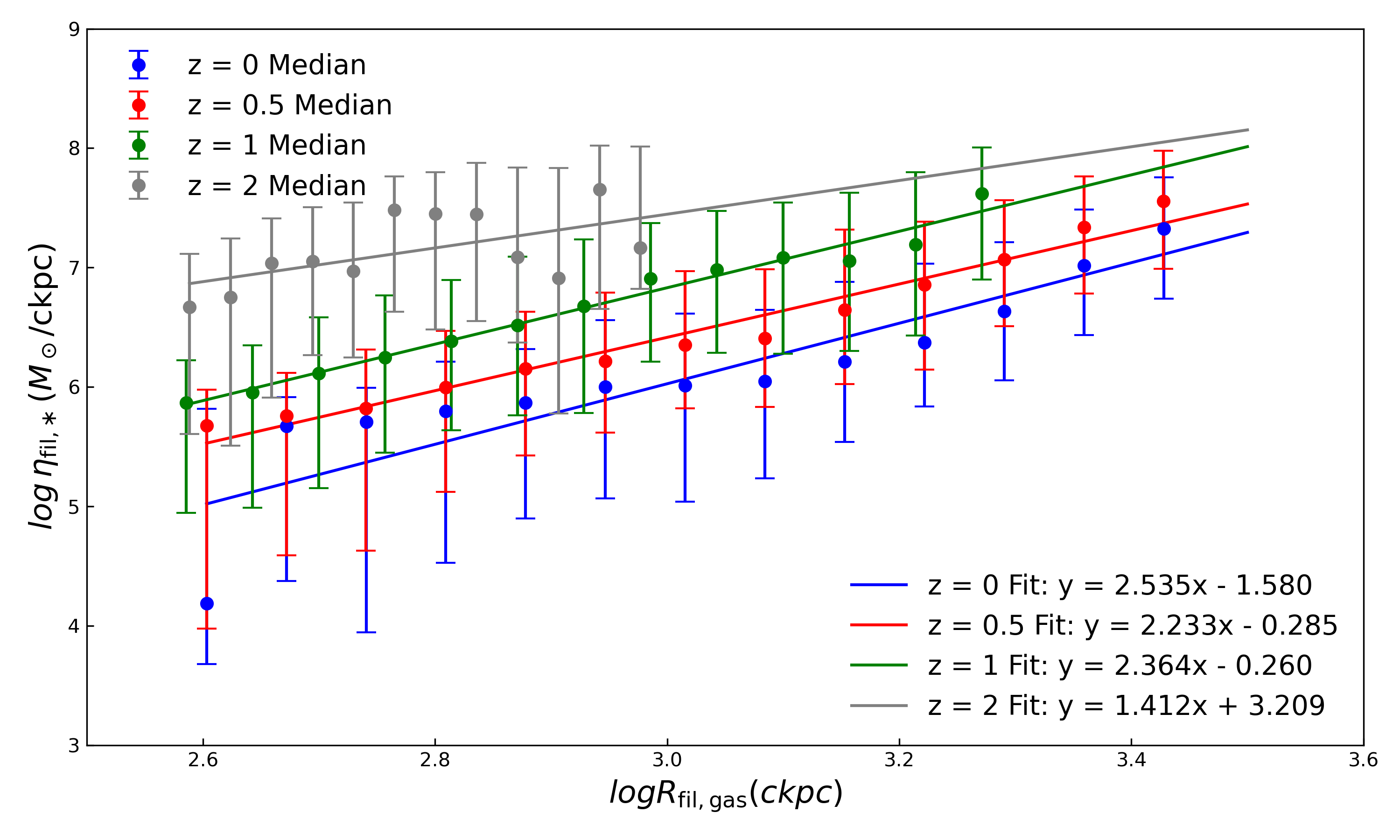}
\caption{Top(bottom): The linear halo (stellar) mass density and segment width relation at different redshift in TNG100. The left and right panels show the results of width derived from dark matter and gas density profiles, respectively. Diverse colors represent results at different redshifts.}
\label{fig:halomass_r_z}
\end{centering}
\end{figure*}

\begin{figure*}[htb]
\begin{centering}
\hspace{-0.0cm}
\includegraphics[width=1.0\textwidth, clip]{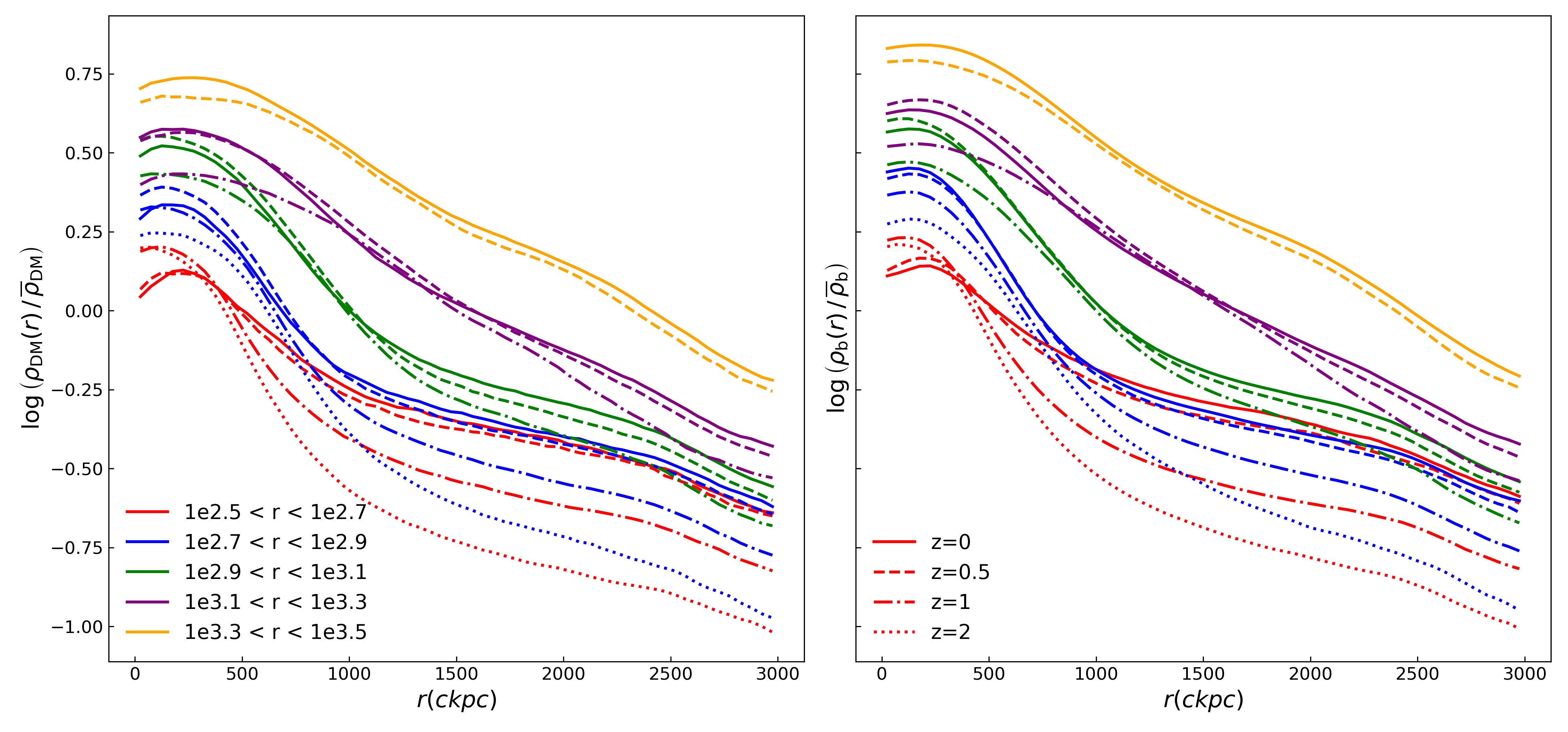}
 \includegraphics[width=1.0\textwidth, clip]{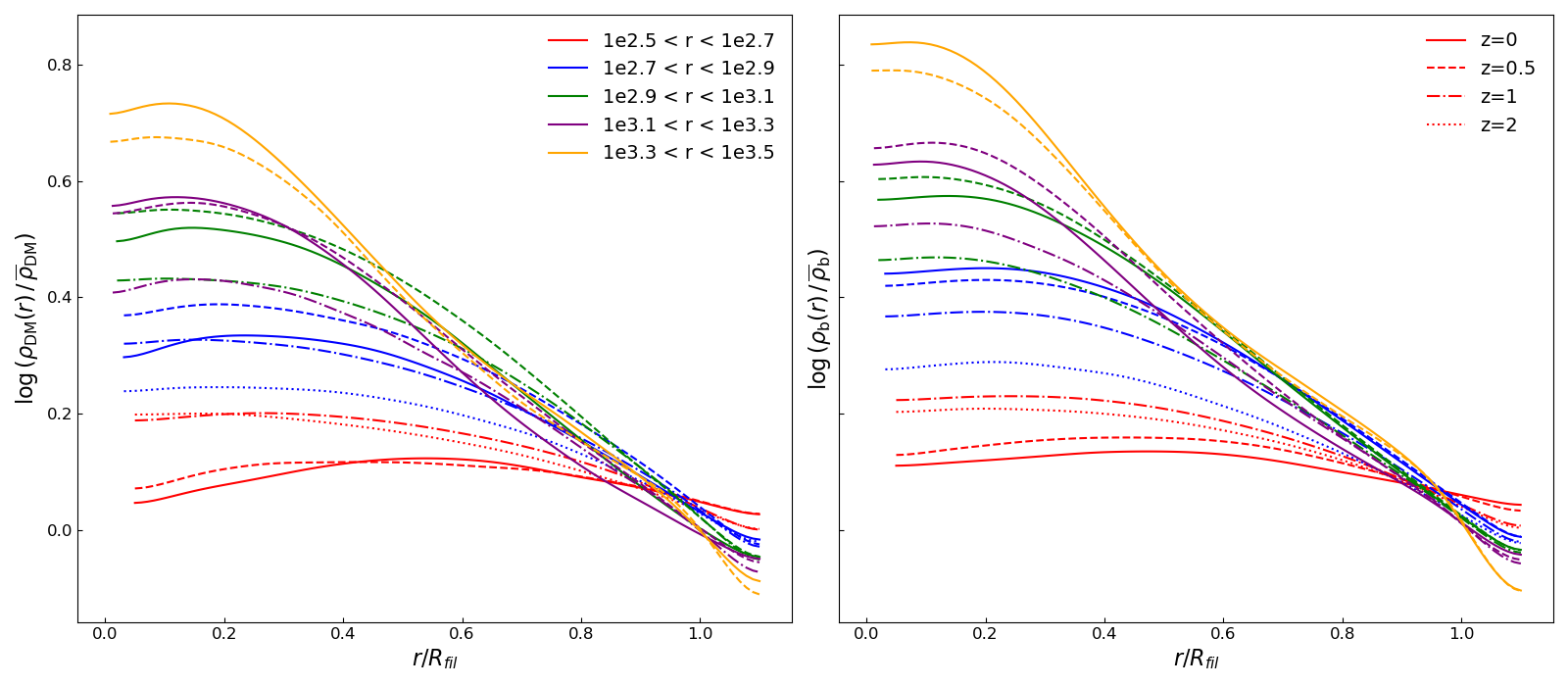}    
\caption{The upper (bottom) row shows the (rescaled) mean density profiles as a function of distance from the spine for filament segments within various width bins in TNG100 at different redshifts. The left and right column show results of dark matter and gas, respectively.}
\label{fig:denprof_tng100_z}
\end{centering}
\end{figure*}

\begin{figure}[htbp]
\begin{centering}
\hspace{-0.0cm}
\includegraphics[width=0.5\textwidth, clip]{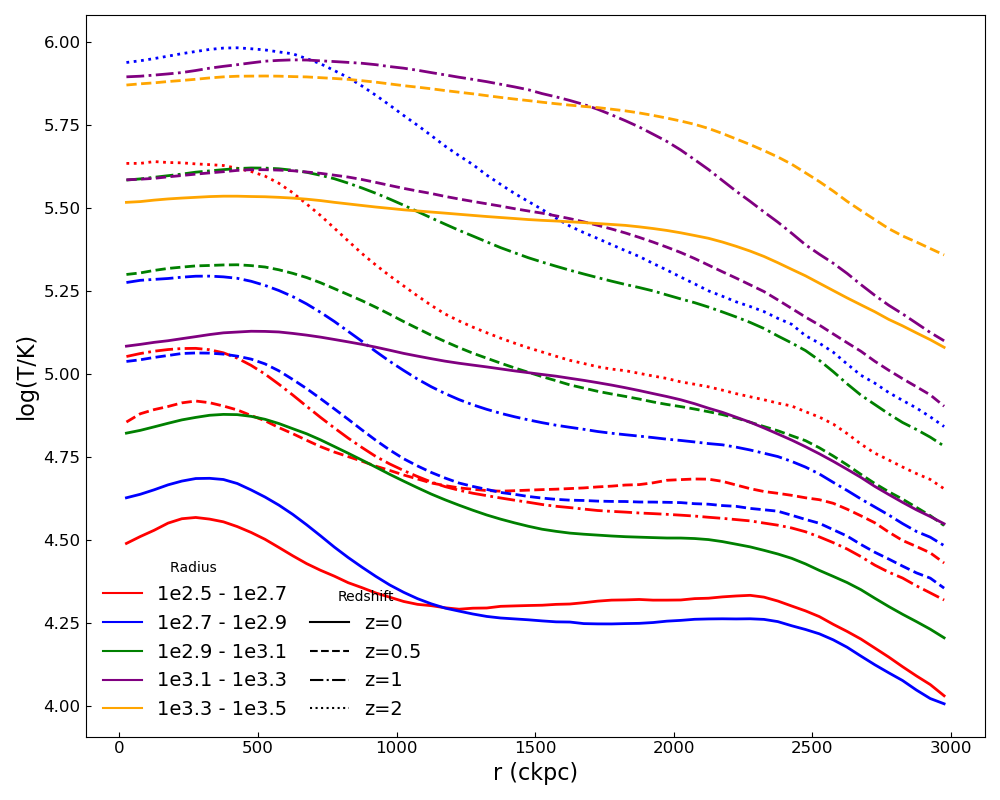}
\caption{The mean temperature profiles of filaments with different width (from $10^{2.5}kpc$ to $10^{3.5}kpc$) at different redshifts in TNG100. }
\label{fig:T_profile_z}
\end{centering}
\end{figure}

\section{Results at higher redshift}

We also examine the filament width distribution in the TNG100 simulation at higher redshifts. Figure~\ref{fig:width_pdf_z} shows that typical filament widths increase over time, with most segments at $z=2$ being thinner than $0.5\,\mathrm{Mpc}$. As redshift decreases, filaments grow through accretion and mergers, with the peak width reaching $\sim 0.3$, $0.6$, $0.75$, and $1.0\,\mathrm{Mpc}$ at $z=2.0$, $1.0$, $0.5$, and $0.0$, respectively. This evolution is consistent with hierarchical structure formation in the $\Lambda\mathrm{CDM}$ model and agrees with findings from \citet{2014MNRAS.441.2923C}, \citet{2021ApJ...920....2Z}, and \citet{2025arXiv250206484B}.

Figure~\ref{fig:halomass_r_z} shows the linear halo (top row) and stellar (bottom row) mass densities versus filament segment width in TNG100 from $z=2$ to $z=0$. Different colors denote different redshifts, and widths are measured using dark matter (left panels) and baryonic matter (right panels). The linear correlations between linear mass density and segment width persists across redshifts. However, filaments with the same width at higher redshift tend to connect higher-density peaks, generally host more halo, particularly those with masses above $10^{12}\,\mathrm{M_\odot}$, resulting in higher linear densities. 

The top row of Figure~\ref{fig:denprof_tng100_z} shows the density profiles of filament segments in TNG100 from $z=2$ to $z=0$. For widths greater than $0.8\,\mathrm{Mpc}$, the profiles evolve only slightly with redshift, which is consistent with \cite{2025arXiv250206484B}. The rescaled density profiles in the bottom row of Figure~\ref{fig:denprof_tng100_z} further demonstrate that the self-similar behavior persists at high redshift. Figure~\ref{fig:T_profile_z} presents the temperature profiles over the same redshift range. While the shapes of the temperature profiles remain similar across redshifts for segments of the same width, the typical temperatures are higher at earlier times, likely due to the greater masses per unit length of these structures.

\section{Discussion}

Our results show good overall consistency among TNG50, TNG100, and TNG300. Moreover, they align well with previous studies, such as \citet{2021ApJ...920....2Z} and \citet{2024ApJ...967..141Z}, which employed simulations run with the RAMSES code \citep{2021ApJ...906...95Z}. The general characteristics of the density and temperature profiles also agree with the recent independent analysis by \citet{2025arXiv250206484B}, which used both the EAGLE and TNG100 simulations and was submitted during the preparation of this work. This consistency across simulation codes and baryonic models supports the robustness of our findings. Additionally, our study highlights the importance of analyzing filaments by dividing them into segments, properties like width and density profiles can vary substantially along their spines.

The number of filament segments identified across different simulations scales approximately linearly with simulation volume. However, mild deviations from this trend, particularly in TNG300, arise due to factors such as cosmic variance, differences in the segmentation procedure, and reduced star formation efficiency at lower resolution. Using a lower stellar mass threshold for galaxy selection in TNG300 could potentially reveal more thin filament segments, slightly influencing the width distribution. Nevertheless, this adjustment is expected to have only a minor impact on other key results, including the relationship between segment width and linear halo/stellar density, as well as the density and temperature profiles, since our analysis already includes a substantial number of segments within the width range of $0.3-3.0$ Mpc.

One of the key findings of our study is the self-similarity of density profiles, normalized by filament width, across segments of varying widths, a result that is also reported in earlier work based on different simulations and baryonic physics \citep[e.g.,][]{2021ApJ...920....2Z}. This self-similarity originate from the gravitational collapse and virialization processes that govern the formation of filaments segment with various width. Furthermore, we find that the rescaled density profiles are well described by a generalized isothermal beta model, consistent with previous studies \citep[e.g.,][]{2021A&A...646A.156T,2021ApJ...920....2Z,2022A&A...661A.115G}.

Our findings on the correlation between filament width and linear halo or stellar density offer a practical tool for estimating filament segment widths, which can be applied to filaments identified from galaxy distributions. Specifically, filaments can first be identified using tools such as DisPerSE, then divided into segments. For each segment, the linear stellar mass density can be calculated from galaxy samples, allowing the filament width to be estimated via the scaling relation established in this work.

These results can help to locating the universe’s missing baryons and improving estimates of the baryon content in cosmic filaments. Theoretical studies suggest that a significant fraction of the missing baryons reside in cosmic filaments. Observational efforts using stacked X-ray and CMB measurements, as well as cross-correlations with galaxy distributions, have reported detections of warm-hot baryons in filaments \citep[e.g.,][]{2019MNRAS.483..223T,2019A&A...624A..48D,2020A&A...643L...2T}. However, these studies often rely on constant or simplified assumptions about filament width, density, and temperature profiles when interpreting signals and estimating baryon content, which introducing notable uncertainties. Our results can help to find the optimal filament samples for locating warm-hot baryons, and to reduce the relevant uncertainties by providing more accurate estimates of filament widths and by offering a detailed understanding of the density and temperature profiles across segments with varying widths. We plan to pursue related investigations in future work.

\section{Conclusions }
\label{sec:conclusions}
We have investigated the widths, density profiles, and temperature distributions of filament segments in the IllustrisTNG simulations, focusing on those with widths between approximately 0.3 and 3.0 Mpc. Our key findings are summarized below.

\begin{itemize}
\item The typical width of filament segments increases gradually with decreasing redshift, consistent with the hierarchical structure formation in the $\Lambda$CDM universe. At $z=2$, most segments have widths below $0.5\,\mathrm{Mpc}$, while by $z=0$, they span $0.4$–$2.75\,\mathrm{Mpc}$, with a median width of approximately $1.3-1.5\,\mathrm{Mpc}$.

\item The linear halo mass density of cosmic filament segments correlates approximately linearly with their width, in broad agreement with \citet{2024ApJ...967..141Z}

\item The linear stellar mass density also shows a linear correlation with filament segment width, suggesting it can serve as a proxy for estimating filament width from observed galaxy distributions.

\item The density profiles as a function of
radius normalized by the segment width of both dark matter and baryons exhibit strong self-similarity across a wide range of filament widths when measured relative to the mass spine. These profiles can be well described by a common functional form similar to the isothermal $\beta$-model. For filaments of a given width, the rescaled density profiles evolve only mildly from $z=2$ to $z=0$.

\item Within the filament width, the gas temperature decreases slowly from the center to the boundary. Thicker filaments generally contain hotter gas than thinner ones. For segments between $0.3$ and $0.5\,\mathrm{Mpc}$, the typical temperature is around $10^{4.5}\,\mathrm{K}$, rising to approximately $10^{5.5}\,\mathrm{K}$ for widths of $2.0$–$3.0\,\mathrm{Mpc}$. This trend is consistent across redshifts. 
\end{itemize}

Results in this study are generally consistent studies using samples based on simulation run by RAMESE(\citealt{2021ApJ...920....2Z}), and EAGLE(\citealt{2025arXiv250206484B}), which suggest that our findings are robust. Our analysis provides a framework for estimating filament widths out to the radius where the density approaches the cosmic mean. This methodology can be applied to filaments identified in observed galaxy distributions. Additionally, our results on density and temperature profiles offer useful insights for locating and estimating the universe’s missing baryons and for understanding the potential influence of filaments on galaxy properties (\citealt{2025arXiv250401245Y}).

\begin{acknowledgments}
We are grateful to the anonymous reviewer for his/her insightful comments and suggestions. We thank Zi-Yong Wu, Xi-Chang Ou-yang for their helpful discussions. This work is supported by the National Natural Science Foundation of China (NFSC) through grant 11733010, 12173102 and 12203107, and by the China Manned Space Program through its Space Application System. The calculation carried out in this work was completed on the HPC facility of the School of Physics and Astronomy, Sun, Yat-Sen University.
\\

\end{acknowledgments}

\begin{contribution}
WSZ conceived the initial research concept and was responsible for writing and submitting the manuscript. QRY conducted the formal analysis and validation and also contributed to the initial draft. GYY assisted with the analysis and validation. JFM, YZ, and LLF contributed to both the analysis and the development of the research concept.


\end{contribution}

%



\software{DisPerSe \citep{sousbie2011persistent}
         }

\vspace{15mm}


\appendix

Each filament identified by \texttt{DisPerSe} is composed of multiple short segments, connected by a series of discrete points, with typical lengths of about 1 Mpc or less. To divide our filament samples from IllustrisTNG simulations into longer, relatively straight segments ranging from $1.5\,\mathrm{Mpc}/h$ to $2.5\,\mathrm{Mpc}/h$, we follow the procedure outlined below:

\begin{enumerate}
    \item For each filament, we start at one of its two terminal nodes and sequentially follow the segment points (discrete sampling points along the filament skeleton, such as saddle-2 points or elements with a critical index of $–1$) identified by \texttt{DisPerSE}. At each segment point, we calculate the angle, $\theta_{adj}$, between two consecutive short segments to quantify local curvature. Simultaneously, we track the cumulative distance traveled from the starting node along the filament.
    
    \item For strongly curved filaments, those containing at least one adjacent segment angle $\theta_{adj} \geq 45^\circ$, we divide the filament at the segment points where this angular threshold is exceeded. After this initial split, we retain the resulting segments with lengths between $1.5\,\mathrm{Mpc}/h$ and $2.5\,\mathrm{Mpc}/h$. Segments longer than $2.5\,\mathrm{Mpc}/h$ are further divided into smaller sub-segments within this target length range. Segments shorter than $1.5\,\mathrm{Mpc}/h$ are excluded from subsequent analysis, as their limited extent often leads to significant fluctuations in density profiles, making it challenging to derive reliable width measurements.
    
    \item For relatively straight filaments, those in which the angles between adjacent segments, as well as the average segment angle, remain consistently below $45^\circ$ along the entire filament, we segment them based on cumulative length. Specifically, we break the filament at selected segment points to ensure that the resulting segments fall within the target range of $1.5\,\mathrm{Mpc}/h$ and $2.5\,\mathrm{Mpc}/h$. Any remaining segments shorter than $1.5\,\mathrm{Mpc}/h$ are also discarded due the reason described above.  

\end{enumerate}

Each filament segment in our sample is composed of multiple initially identified segments from DisPerSE.

\begin{figure}[htbp]
\begin{centering}
\hspace{-0.0cm}
\includegraphics[width=0.5\textwidth, clip]{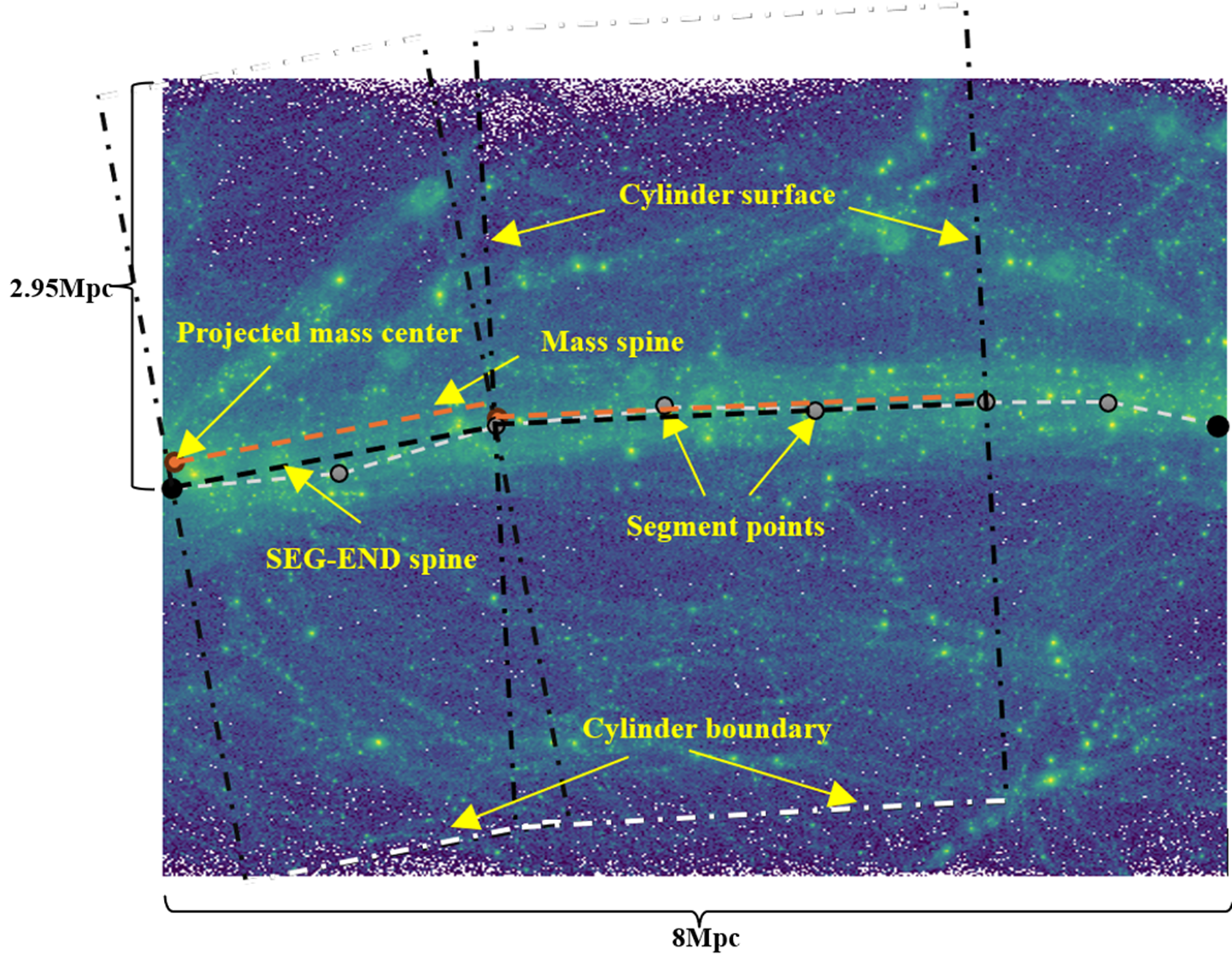}
\caption{Illustration of the two definitions of segment spines. The blue-green background shows the dark matter distribution. Solid black dots mark the two ends of the filament, while gray dots represent the segment (sampling) points. The filament is divided into several segments. Orange and black dashed lines indicate the mass spine and the SEG-END spine, respectively, for the left and middle segments.}
\label{fig:two_spines}
\end{centering}
\end{figure}

Figure \ref{fig:two_spines} illustrates the two definitions of segment spines described in section 2.3.

\bibliography{main}
\bibliographystyle{aasjournalv7}



\end{document}